\documentclass[11pt,a4paper]{article}
\usepackage[utf8]{inputenc}
\usepackage{authblk}
\usepackage{fullpage}
\usepackage{microtype}
\usepackage{multirow}
\usepackage{xspace}
\usepackage{todonotes}

\usepackage{amsfonts}
\usepackage{hyperref}
\usepackage{url}

\usepackage{caption}
\usepackage{subcaption}
\usepackage{graphicx}
\usepackage{tikz}
\usetikzlibrary{quantikz}

\newcommand{\ie}{\emph{i.e.}\xspace}

\newcommand{\figref}[1]{Figure~\ref{#1}}
\renewcommand{\ket}[1]{|#1\rangle}

\title{Enhancing variational quantum state diagonalization using reinforcement learning techniques}

\author{Akash Kundu$^{1,2\footnote{Corresponding author, \href{mailto:akundu@iitis.pl}{\texttt{akundu@iitis.pl}}}}$, Przemys\l{}aw Bede\l{}ek$^3$, Mateusz Ostaszewski$^3$, Onur Danaci$^{4,5}$, Yash~J.~Patel$^{6}$, Vedran~Dunjko$^{4,6}$, Jaros\l{}aw A. Miszczak$^1$}
\affil{$^1$Institute of Theoretical and Applied Informatics, Polish Academy of Sciences, Gliwice, Poland}
\affil{$^2$Joint Doctoral School, Silesian University of Technology, Gliwice, Poland}
\affil{$^3$Warsaw University of Technology, Institute of Computer Science, Warsaw, Poland}
\affil{$^4$Applied Quantum Algorithms, Lorentz Institute, Leiden University, Leiden, the Netherlands}
\affil{$^5$Leiden Institute of Physics, Leiden University, Leiden, the Netherlands}
\affil{$^6$Leiden Institute of Advanced Computer Science, Leiden University, Leiden, the Netherlands}
\date{}

\begin{document}
\maketitle
\begin{abstract}
The variational quantum algorithms are crucial for the application of NISQ computers. Such algorithms require short quantum circuits, which are more amenable to implementation on near-term hardware, and many such methods have been developed. One of particular interest is the so-called variational quantum state diagonalization method, which constitutes an important algorithmic subroutine and can be used directly to work with data encoded in quantum states. In particular,  it can be applied to discern the features of quantum states, such as entanglement properties of a system, or in quantum machine learning algorithms. In this work, we tackle the problem of designing a very shallow quantum circuit, required in the quantum state diagonalization task, by utilizing reinforcement learning (RL). We use a novel encoding method for the RL-state, a dense reward function, and an $\epsilon$-greedy policy to achieve this. We demonstrate that the circuits proposed by the reinforcement learning methods are shallower than the standard variational quantum state diagonalization algorithm and thus can be used in situations where hardware capabilities limit the depth of quantum circuits. The methods we propose in the paper can be readily adapted to address a wide range of variational quantum algorithms.
\end{abstract}

\section{Introduction}

In the last few decades, researchers from various scientific disciplines have come together to study and develop quantum algorithms and their experimental realization. Among the originally proposed quantum algorithms, many require millions of physical qubits to be implemented on quantum hardware to deal with instance sizes of real-world importance. Unfortunately, the existing quantum hardware is limited to the order of a few hundred physical qubits, and these are called Noisy Intermediate-Scale Quantum (NISQ) devices. The NISQ algorithms are small and prone to noise and decoherence, and thus, one needs to consider Variational Quantum Algorithms (VQAs) that can work under such restrictions.

Among the class of VQAs, Variational Quantum State Diagonalization (VQSD)~\cite{larose2019variational} is an algorithm that utilizes a quantum-classical hybrid procedure to identify the unitary rotation under which a given quantum state becomes diagonal in the computational basis, \ie it diagonalizes a quantum state. It has several applications, including quantum state fidelity estimation~\cite{cerezo2020vqfe}, device certification~\cite{kundu2022variational}, Hamiltonian diagonalization~\cite{zeng2021variational}, and as a method to extract entanglement properties of a system \cite{larose2019variational, cerezo2022variational}. VQSD generalizes the well-studied problem of quantum state preparation, which can be understood as quantum state tomography for pure states\footnote{{If one is given a quantum state $\ket{\psi}$, then VQSD can potentially find a short-depth circuit that approximately prepares $\ket{\psi}$.}}. Considering it has applications that range from quantum information to condensed matter physics, an efficient way to deal with quantum state diagonalization may lead to interesting insights in these fields.

We note that there exist algorithmic exact methods for quantum state diagonalization based on quantum principal component analysis (qPCA)~\cite{lloyd2014quantum}. However, they lead to deeper circuits that could, in principle, be obtained with variational methods. However, to achieve this, the most challenging aspect of VQSD is to construct an \emph{efficient} ansatz (which refers to the unitary) that diagonalizes a given quantum state. For the analysis in this paper, we consider the following factors as the indicators for \emph{ansatz efficiency}: (1) the depth, understood as the number of parallel operations in the ansatz; (2) the total number of quantum gates, and (3) the accuracy in the estimation of eigenvalues.

In the standard VQSD methods \cite{larose2019variational, cerezo2022variational}, a Layered Hardware Efficient Ansatz (LHEA) is utilized. A single layer of the ansatz contains two-qubit gates acting on neighbouring qubits. Although the LHEA parameter count increases linearly with the number of layers and qubits, it has trainability issues and often encounters local minima~\cite{larose2019variational}. To tackle the trainability issue, instead of using a fixed structure of LHEA, the authors allow additional updates (\ie changes in the ansatz structure) during the classical optimization process~\cite{larose2019variational}. In this process, every optimization step minimizes the cost function with a small random change to the ansatz structure. The new structure is approved or rejected based on a simulated annealing scheme~\cite{cincio2018learning}. Although the varying structure LHEA outperforms fixed structure LHEA, the number of gates in the quantum circuit increases rapidly as we scale the size of the quantum state. Hence, the problem of finding a method to construct an ansatz that satisfies all efficiency criteria is still an open problem.

In the case of some VQAs, to address the challenges of finding the architecture of ansatz, methods have been introduced that draw on the insight and techniques of machine learning~\cite{ostaszewski2021reinforcement, kuo2021quantum, ye2021quantum, he2023gnn, bolens2021reinforcement}, such as a process of automating the architecture engineering of quantum circuits is known as Quantum Architecture Search (QAS)~\cite{kuo2021quantum, zhang2022differentiable, du2022quantum}. Recent studies have strongly suggested that Double deep Q-networks (DDQN) in Reinforcement Learning (RL) can successfully solve QAS problems~\cite{ostaszewski2021reinforcement, ye2021quantum}, performance improvement in QAOA variants~\cite{patel2022reinforcement} as well as the task of quantum compiling~\cite{moro2021quantum}.

\paragraph*{Contributions}
Following the above line of work, we introduce a Reinforcement Learning (RL) driven VQSD method (\ie, RL-VQSD), {which automates the search for optimal succinct ansatz (\ie RL-ansatz). The RL-VQSD algorithm constitutes: 
\begin{enumerate}
    \item A novel depth-based binary encoding scheme \cite{rl-vqe-paper} to encode the RL-state.
    \item A dense reward function, which we introduce in the paper crafted particularly for the task of quantum state diagonalization.
    \item A Double deep Q-network (DDQN) with an $\epsilon$-greedy policy for better stability. 
\end{enumerate}
Using these components we demonstrate that the ansatz proposed by the RL-agent can successfully diagonalize arbitrary mixed quantum states of full-rank with a smaller number of gates and depth compared to the existing ansatz structures}. We exemplify the functioning of the RL-VQSD by diagonalizing the quantum states arising in condensed matter physics {while maintaining a short depth and gate count} of the resulting RL-ansatz. {Moreover, a deeper investigation reveals that the combination of the binary encoding of the RL-state and the dense reward function is responsible for the success of diagonalizing larger quantum states.}
Finally, we demonstrate the hardness of the problems by utilizing a random agent in the VQSD algorithm and show the performance of the random agent significantly decreases as we scale up the qubits in the quantum state. Moreover, we show that the RL-agent not only provides us with a more consistent outcome, but it gives significantly better circuit depth, gate count, and approximation quality compared to the random agent.

The rest of this paper is organized as follows. In Section~\ref{sec:preliminaries}, we review the standard methods for variational quantum state diagonalization and provide an overview of the ansatz construction, and reinforcement learning. In Section~\ref{sec:methods}, we describe the proposed scheme for the construction of variational quantum state diagonalization circuits, including the method for encoding quantum circuits, the dense reward function, and the performance comparison of the encoding and the reward. Section~\ref{sec:results} summarises the numerical results obtained to demonstrate the application of the proposed RL-VQSD. Finally, in Section~\ref{sec:final}, we briefly summarize the contribution and provide some remarks concerning the possible extension of the introduced approach.

\section{Preliminaries}\label{sec:preliminaries}

This section briefly reviews the standard methods of variational quantum state diagonalization. We also outline the standard procedure of constructing an ansatz and introduce basic concepts from reinforcement learning.

\subsection{Variational quantum state diagonalization}

Classical methods for diagonalization typically scale polynomially with the dimension of the matrix~\cite{demmel2008performance}. Similarly, the number of measurements required for quantum state tomography scales polynomially with the dimension of the Hilbert space. Moreover, as discussed, the qPCA is costly to implement in NISQ devices.

\begin{figure}[t!]
\centering
\includegraphics[width = 0.75\linewidth]{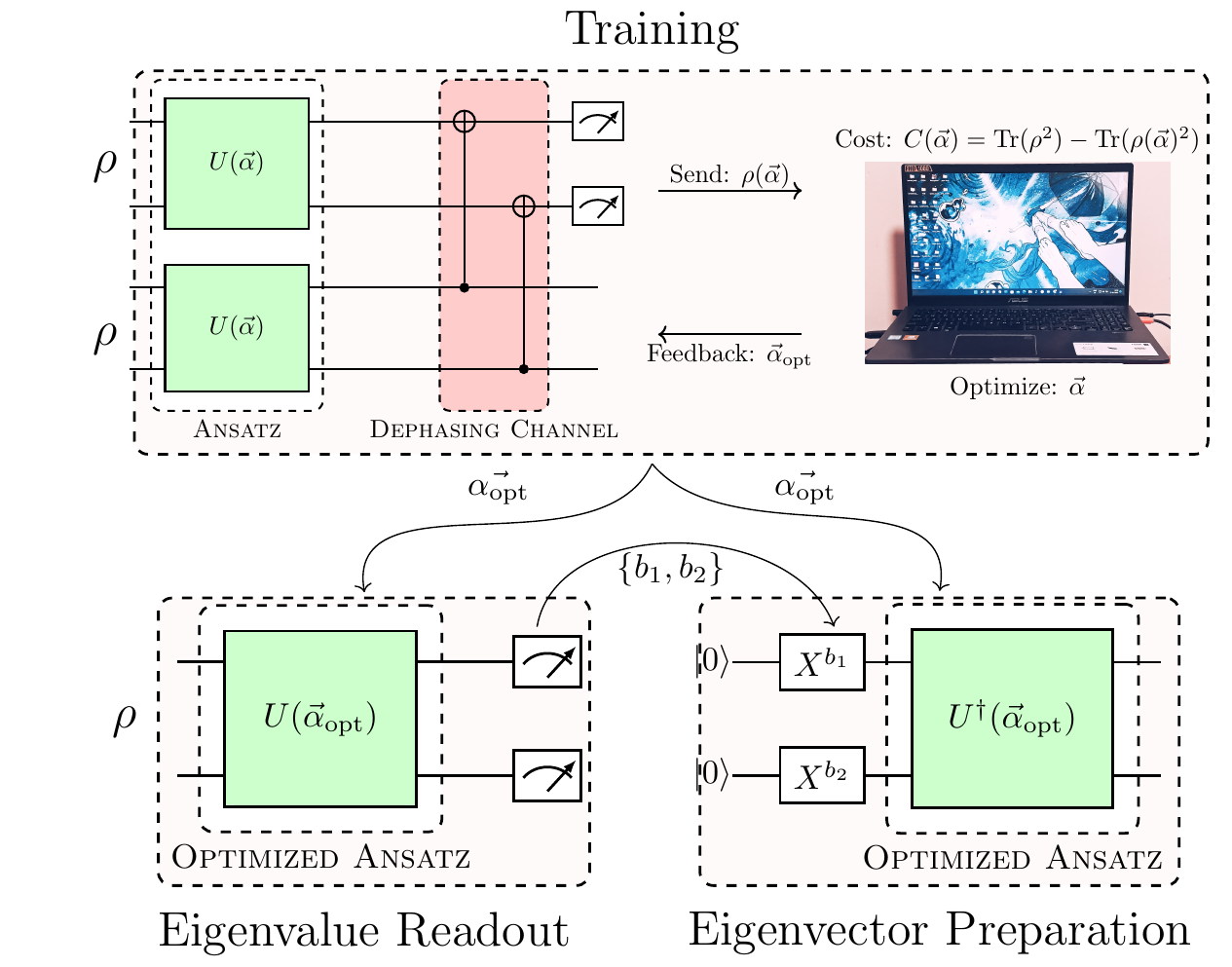}
\caption{\textbf{Elements of Variational Quantum State Diagonalization (VQSD) algorithm}. In the presented example, we consider the diagonalization for the $2$-qubit input state. It should be noted that to diagonalize the $N$ qubit quantum state the algorithm requires $2N$ number of qubits in the algorithm.}
\label{fig:ill-vqsd}
\end{figure}

To tackle these issues, a hybrid quantum-classical method for quantum state diagonalization -- Variational Quantum State Diagonalization (VQSD) -- has been proposed in~\cite{larose2019variational}. For a quantum state $\rho$, the algorithm is composed of three subroutines:
\begin{itemize}

	\item \textsc{Training} In 
 this subroutine, for a given state 
 $\rho$, one optimizes the 
 parameters $\vec{\theta}$ of a 
 quantum gate sequence 
 $U(\vec{\theta})$, which 
 (ideally) after optimization 
 satisfies 
	\begin{equation}
		\rho' = U(\vec{\theta}_\text{opt})\rho U(\vec{\theta}_\text{opt})^\dagger = \rho_\text{diag},
	\end{equation}
	where $\rho_\text{diag}$ is the diagonalized $\rho$ in its eigenbasis and $\vec{\theta}_\text{opt}$ are the optimal angles. One can utilize classical gradient-based methods such as SPSA and Gradient-Descent or gradient-free optimization methods such as COBYLA~\cite{powell1994direct} and POWELL~\cite{powell2006fast} in the training process.
 
    \item \textsc{Eigenvalue Readout} In this subroutine, using the optimized unitary $U(\vec{\theta}_\text{opt})$ and one copy of state $\rho$, one can extract -- for low-rank states -- all the eigenvalues or --  for full-rank state -- the largest eigenvalues. This is achieved by measuring the ${\rho}'$ in the computational basis, $\mathbf{b}=b_1b_2\ldots b_n$, as follows
    \begin{equation}
        \lambda' = \bra{\mathbf{b}}\rho'\ket{\mathbf{b}},
    \end{equation}
    where $\lambda'$ are inferred eigenvalues.
    
    \item \textsc{Eigenvector Preparation} In the final {step}, one can prepare the eigenvectors associated with the largest eigenvalues. If $\mathbf{b}'$ is a bit string associated with $\lambda'$ then one can get the inferred eigenvectors $|v_{\mathbf{b}'}'\rangle$ as follows
    \begin{equation}
        |{v}'_{{\mathbf{b}}'}\rangle = U(\theta_\textrm{opt})^\dagger\ket{{\mathbf{b}}'} = U(\theta_\textrm{opt})^\dagger\left(X^{b_1}\otimes\ldots\otimes X^{b_n} \right)\ket{\mathbf{0}}.
    \end{equation}
\end{itemize}
The workflow in the VQSD procedure is illustrated in \figref{fig:ill-vqsd}.

The cost function proposed in~\cite{larose2019variational} as a part of the training process is a function of the \textit{purity} of the state that needs to be diagonalized.  It takes the following form
\begin{equation}
	C(\vec{\theta}) = \text{Tr}(\rho^2) - \text{Tr}(\mathcal{D}({\rho}')^2),\label{eq:cost_func}
\end{equation}
where $\mathcal{D}$ denotes a dephasing channel, that eliminates the off-diagonal elements. When $C(\vec{\theta})$ is sufficiently close to zero, one can say that the quantum state is diagonalized. It should be noted that there are many ways to define a cost function that quantifies how far ${\rho}'$ is from being diagonal~\cite{baumgratz2014quantifying}. However, due to computational purposes, we choose the cost function of the form given in Eq.~\ref{eq:cost_func}.

\subsection{Ansatz construction}
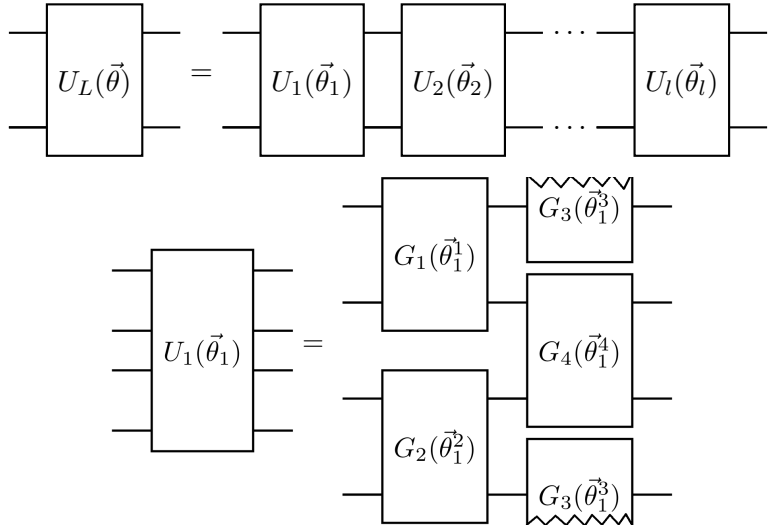
\begin{figure}[h!]
    \centering
    \begin{tikzpicture}
    \node[scale=1] {
        \begin{quantikz}
		& \gate[wires=2][1cm]{U_L(\vec{\theta})} & \qw \\
		& \qw &\qw
	\end{quantikz}
	=\begin{quantikz}
		& \gate[wires=2][1cm]{U_1(\vec{\theta}_1)} & \gate[wires=2][1cm]{U_2(\vec{\theta}_2)} & \ \ldots\ \qw &  \gate[wires=2][1cm]{U_l(\vec{\theta}_l)} & \qw\\
	& \qw & \qw & \ \ldots\ \qw & \qw & \qw
	\end{quantikz}
        };
    \end{tikzpicture}
    \centering
    \includegraphics[width = 0.5\linewidth]{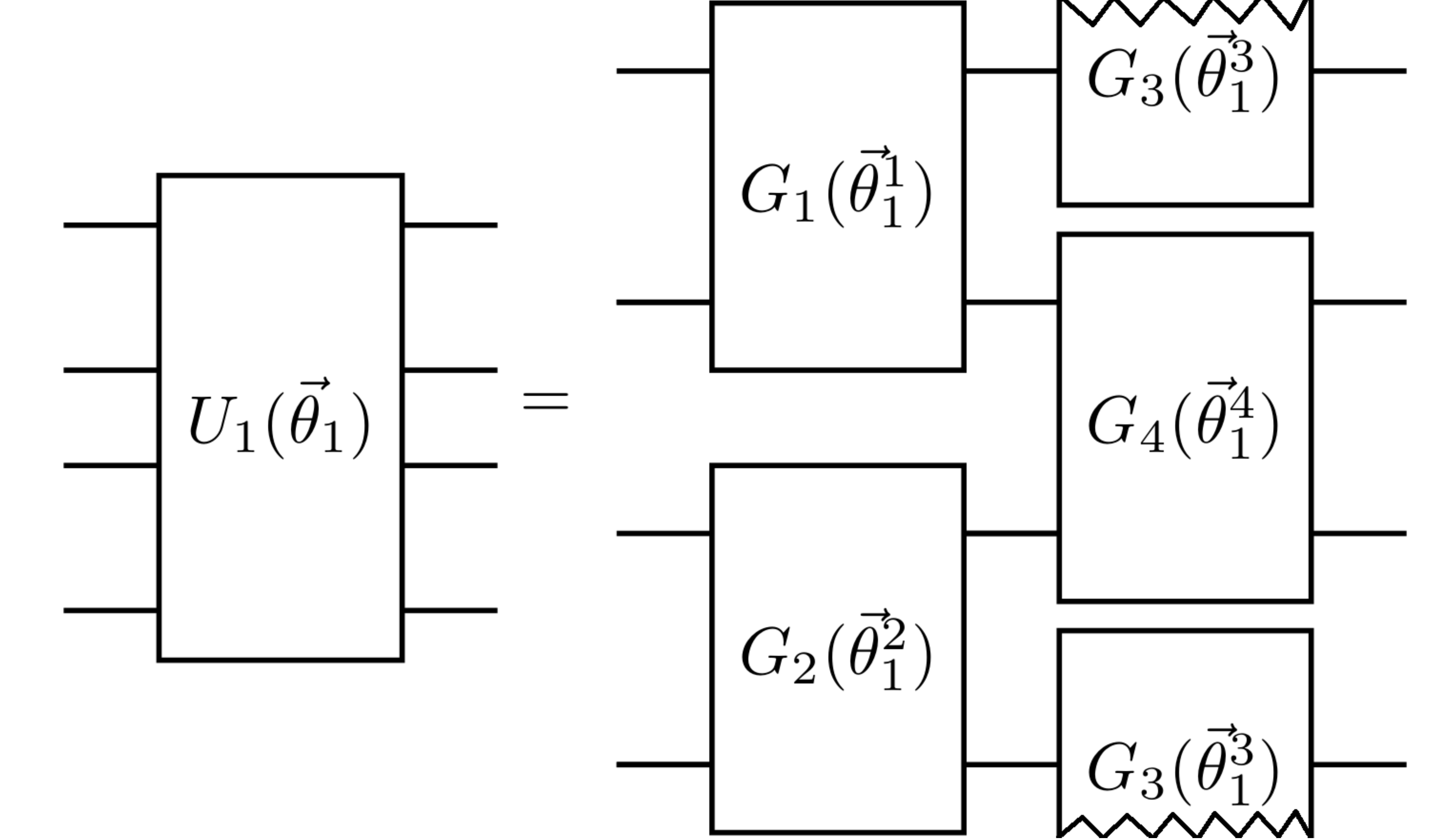}
    \caption{\textbf{Structure of a layered hardware efficient ansatz, where the ansatz $U_l(\vec{\theta})$ is decomposed into layer-wise unitaries $U_l(\vec{\theta}_l)$ for $l = 1,2,\ldots, l$}. Each gate $U_l(\vec{\theta}_l)$ is further decomposed into two-qubit rotations. For $\vec{\theta}_i^j$, index $i$ denotes the layer number, and $j$ is the index specifying the parameter count.}
    \label{fig:layered_ansatz}
\end{figure}
    
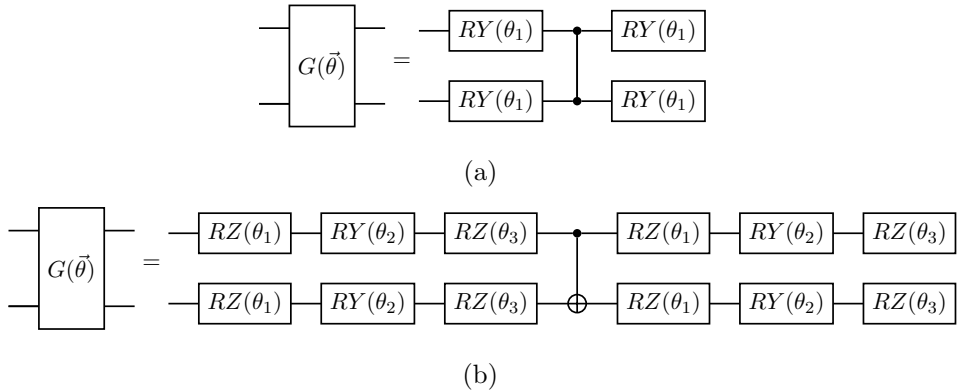
\begin{figure}[h!]
    \begin{subfigure}{\textwidth}
    \centering
    \begin{tikzpicture}
    \node[scale=0.8] {
        \begin{quantikz}
		& \gate[wires=2][1cm]{G(\vec{\theta})} & \qw \\
		& \qw &\qw
	\end{quantikz}
	=\begin{quantikz}
		& \gate{RY(\theta_1)} & \ctrl{1} & \gate{RY(\theta_1)} \\
		& \gate{RY(\theta_1)} & \ctrl{-1} & \gate{RY(\theta_1)}
	\end{quantikz}
        };
    \end{tikzpicture}
    \caption{}
    \label{fig:layered-anzatz-one}
    \end{subfigure}
    \begin{subfigure}{\textwidth}
    \centering
    \begin{tikzpicture}
    \node[scale=0.8] {
        \begin{quantikz}
		& \gate[wires=2][1cm]{G(\vec{\theta})} & \qw \\
		& \qw &\qw
	\end{quantikz}
	=\begin{quantikz}
		& \gate{RZ(\theta_1)} & \gate{RY(\theta_2)} & \gate{RZ(\theta_3)} & \ctrl{1} & \gate{RZ(\theta_1)} & \gate{RY(\theta_2)} & \gate{RZ(\theta_3)}\\
		& \gate{RZ(\theta_1)} & \gate{RY(\theta_2)} & \gate{RZ(\theta_3)} & \targ{} & \gate{RZ(\theta_1)} & \gate{RY(\theta_2)} & \gate{RZ(\theta_3)}
	\end{quantikz}
        };
    \end{tikzpicture}
    \caption{}
    \label{fig:layered-anzatz-three}
    \end{subfigure}
    \caption{\textbf{Two possible decompositions of the two-qubit rotations in each layer-wise unitary $U_i(\vec{\theta}_i)$}. It can be constructed into two forms with (\subref{fig:layered-anzatz-one}) one and (\subref{fig:layered-anzatz-three}) three parameters respectively.}
    \label{fig:two_qubit_rot_decomp}
\end{figure}
    
In the \textsc{Training} subroutine of \figref{fig:ill-vqsd}, the correct choice of the ansatz is crucial, as it is the main factor determining whether the diagonalization task can be performed. Additionally, the choice of the ansatz can also impact the execution of the \textsc{Eigenvalue Readout} and \textsc{Eigenvector Preparation}, as one has to use it in both cases.

In many instances of VQAs, the structure of the ansatz is dictated by the underlying problem. For example, in Variational Quantum Eigensolver (VQE)~\cite{peruzzo2014variational} and the Quantum Approximate Optimization Algorithm (QAOA)~\cite{farhi2014quantum}, the ansatz can be defined based on the problem Hamiltonian. In VQE, the ansatz is constructed through the so-called Unitary Coupled Cluster (UCC)~\cite{ucc3, ucc2, ucc1} method, and in QAOA, it is given by first-order Trotterization of the time-dependent Hamiltonian corresponding to the adiabatic preparation of the ground state. However, this is not the case for the VQSD algorithm; for an arbitrary unknown quantum state, the algorithm has no problem-inspired ansatz. 

In the previous works~\cite{larose2019variational, cerezo2020vqfe} to solve the optimization part, the authors proposed a fixed structure for an ansatz, namely Layered Hardware Efficient Ansatz (LHEA). This type of ansatz is depicted in \figref{fig:layered_ansatz} where each layer $L \in [1..l]$ of $U_L(\vec{\theta})$ consists of a set of optimization parameters $\vec{\theta}\equiv\theta_i^j$, where $i$ denotes the total number of layers and $j$ is the number of parameters per layer. Each layer consists of two-qubit rotation gates which follow a periodic boundary condition. In LHEA, there are two possible ways to construct the two-qubit parameterized gates, which are depicted in \figref{fig:two_qubit_rot_decomp}.

Instead of diagonalizing with a fixed structure ansatz, one can allow it to vary during the optimization process. This scenario starts from a two-qubit parameterized gate on random qubits, and then the gate sequence is optimized by minimizing the cost function and changing the gate-set structure. Hence, the gate sequence is allowed to grow if the algorithm fails to minimize the cost function for a specified number of iterations. Then, one adds an identity gate spanned by new variational parameters that are randomly added to the ansatz. This step is equivalent to adding a layer to the ansatz. This method is discussed in more detail in~\cite{cincio2018learning}.

To address the lack of a definitive structure for diagonalizing unitary, in this paper we utilize reinforcement learning to automate the exploration for an efficient ansatz construction.

\subsection{Reinforcement learning}
{In Reinforcement Learning (RL), an agent interacts with its environment to learn an optimal policy by trial and error approach~\cite{sutton2018reinforcement}. 
An RL process can be modeled as a Markov Decision Process (MDP) defined by the tuple $(S,A,P,R)$, where $S$ and $A$ represent the state and
action spaces, the function $P : S \times S \times A \rightarrow [0, 1]$ defines the transition dynamics, and $R : S \times A \rightarrow \mathbb{R}$ describes the reward function of the environment.
In this work, we consider the action $A$ and the state $S$ to be finite and discrete sets.
An episode describes all interactions between an agent and its environment until a user-specified termination condition is met.}

{An agent's behaviour in the environment is governed by a stochastic policy $\pi(a|s) : S \times A \rightarrow [0, 1]$, for $a \in A$ and $s \in S$.
The metric that assesses an agent's performance is given by the \emph{return} and takes the form of a discounted sum as follows
$$
G(\tau)=\sum_{j=0}^{T-1} r_j \gamma^{j+1},
$$
where $\tau=\left(s_0, a_0, r_0, \ldots, s_{T-1}, a_{T-1}, r_{T-1}\right) \in(S \times A \times \mathbb{R})^T$ is the interaction sequence, $T$ is a fixed length called horizon, and $\gamma$ is an environment-specific discount factor. 
The agent's objective is to determine the optimal policy that maximizes the expected return.}

{In a large unknown environment, the agent needs to be able to adapt to many different situations and develop multiple strategies at the same time. 
Hence, highly expressive function approximators such as deep neural networks to parametrize the agent's policy $\pi$ can be advantageous.}

{Here, we settled on using a Double deep Q-network (DDQN)~\cite{vanhasselt2016deep}.
DDQN is a Q-learning algorithm based on the standard deep Q-network (DQN)~\cite{mnih2015human}, which features two neural networks to increase the stability of the prediction of Q-values for each state and action pair. 
We represent the state space as an ordered list of layers that are composed of a single depth of the quantum circuit.
An action space is defined by a list of four numbers, corresponding to \texttt{RX}, \texttt{RY}, \texttt{RZ}, and \texttt{CNOT} quantum gates.
For the sake of brevity, we defer the detailed description of the DDQN algorithm in Appendix~\ref{app:ddqn}.}

\subsection{Error quantification}
To quantify the eigenvalue error throughout the paper, we use the following {figure of merit}~\cite{larose2019variational}
\begin{equation}
    \Delta_i = \sum_{i=1}^m\left(\lambda_i - \lambda_i'\right)^2,
    \label{eq:statistical_error}
\end{equation}
where $m$ represents the number of the largest eigenvalues, $\lambda_i$ is the true eigenvalue and ${\lambda}_i'$ is the inferred eigenvalue {obtained from the \textsc{eigenvalue readout} subroutine}. In the ideal case, where the state is completely diagonalized, $m = 2^n$ indicates all the eigenvalues have been considered. Throughout the paper, we set $m = 2^n$ if not specified explicitly otherwise.

\section{Proposed approach}\label{sec:methods}
In this section, we give the details of the proposed RL-VQSD as depicted in \figref{fig:rl_vqsd}. {In our modeling of the algorithm, the states of the environment encode the possible architectures of the quantum circuit (\ie, the ansatz), and the actions correspond to a gate. At first, we briefly
\begin{figure}[h!]
    \centering
    \includegraphics[width=0.8\linewidth]{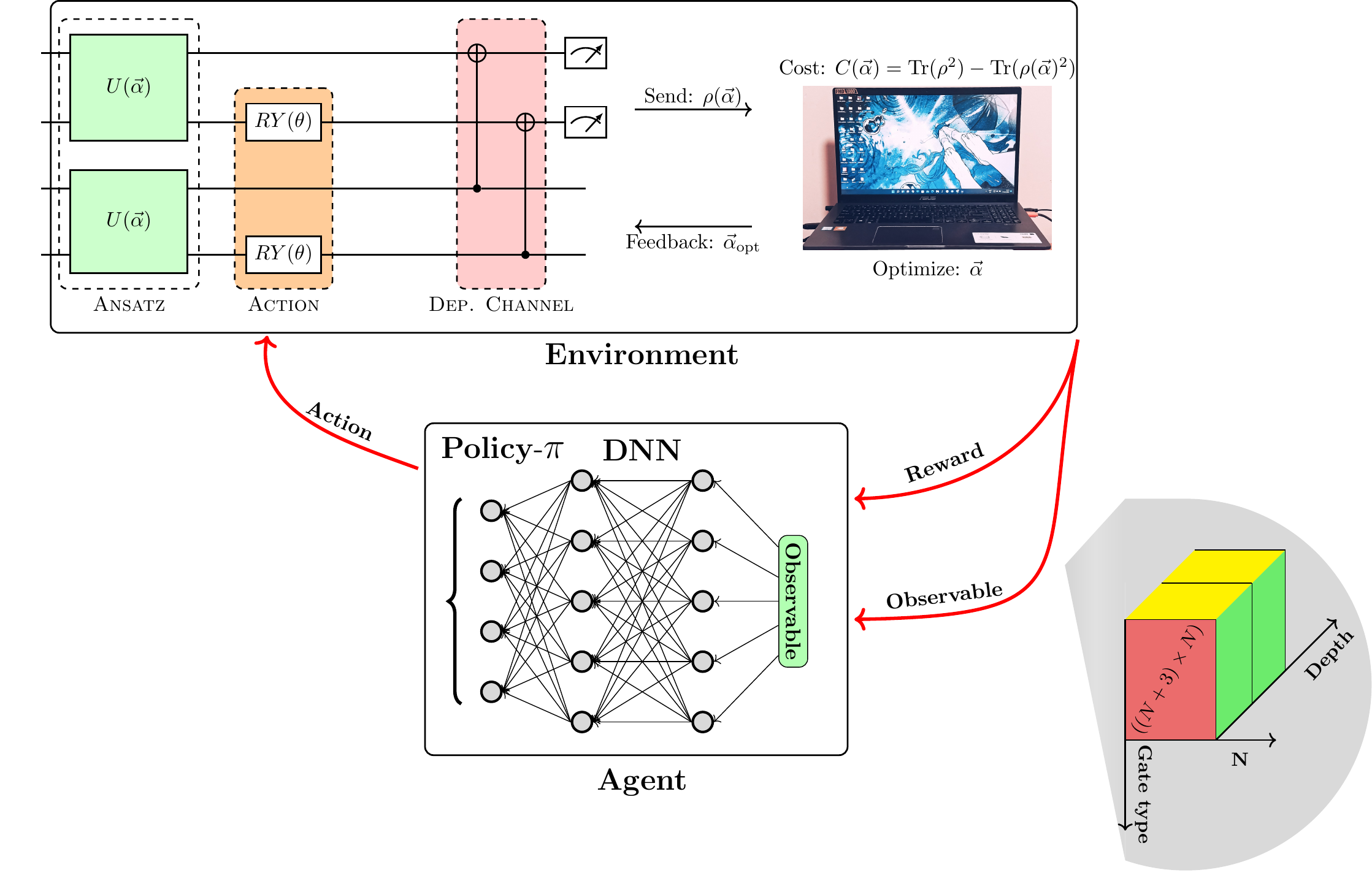}
    \caption{{\textbf{Illustration of the RL-VQSD process}. In this process, the VQA is represented as the environment and the ansatz as the RL-state. The RL-agent receives the optimized cost function in the form of a reward and the RL-state from the environment. Following an $\epsilon$-greedy policy, the agent then decides on an action (i.e., a quantum gate), which in the next step updates the RL-state. Utilizing the new RL-state the VQA optimizes the cost function and generates a new reward function to feed it to the agent. This process is repeated until all the steps in an episode are exhausted, or the cost function reaches a predefined threshold value. Throughout the paper, we start the RL-VQSD with an empty circuit and at each step, the agent chooses an action to construct the RL-ansatz, indicating $U(\vec{\alpha})=\mathbb{I}$.}}
    \label{fig:rl_vqsd}
\end{figure}
discuss the sub-components of the RL-VQSD which include (1) a binary encoding for the RL-state, (2) a one-hot encoding to define actions, and (3) the engineering of a dense reward function. Next, we discuss the agent-environment settings and hyperparameters relevant to RL-VQSD. Finally, we benchmark the performance of the binary encoding scheme and the dense reward function in comparison with the encoding and reward proposed in~\cite{ostaszewski2021reinforcement}, showing that the success of RL-agent is heavily dependent on a well-engineered encoding scheme for RL-state and reward function.

\subsection{Encoding scheme for state}\label{sec:encoding}
Motivated by the ideas in~\cite{ostaszewski2021reinforcement} and~\cite{fosel2021quantum}, in \cite{rl-vqe-paper} a binary encoding scheme was introduced. In this scheme, the gate structure of the ansatz is expressed as a tensor of dimension $\left[D_\textrm{max} \times ((N + 3) \times N )\right]$, where $N$ represents the size of the problem and $D_\textrm{max}$ is the considered maximum depth of the ansatz. For VQSD, $N$ {represents the number of qubits in the quantum state} that need to be diagonalized. The proposed encoding  can be explained through the following two points:
\begin{enumerate}
    \item \textbf{Freedom in connectivity} The encoding enables \textit{all-to-all} qubit connectivity, but it can be restricted by considering \textit{unidirectional nearest neighbour} connections only. In this scenario, the matrix dimension $\left((N+3)\times N \right)$ is reduced to $\left(4\times N \right)$. One should note that in the case of a two-qubit gate, one is not required to keep track of the control and target simultaneously. Hence, defining one argument of the two-qubit gate implicitly provides information about the other argument due to its nearest neighbour and unidirectional nature. A similar encoding scheme is described in~\cite{fosel2021quantum}.
    
\begin{figure}[ht!]
\centering
\includegraphics[width = 0.5\linewidth]{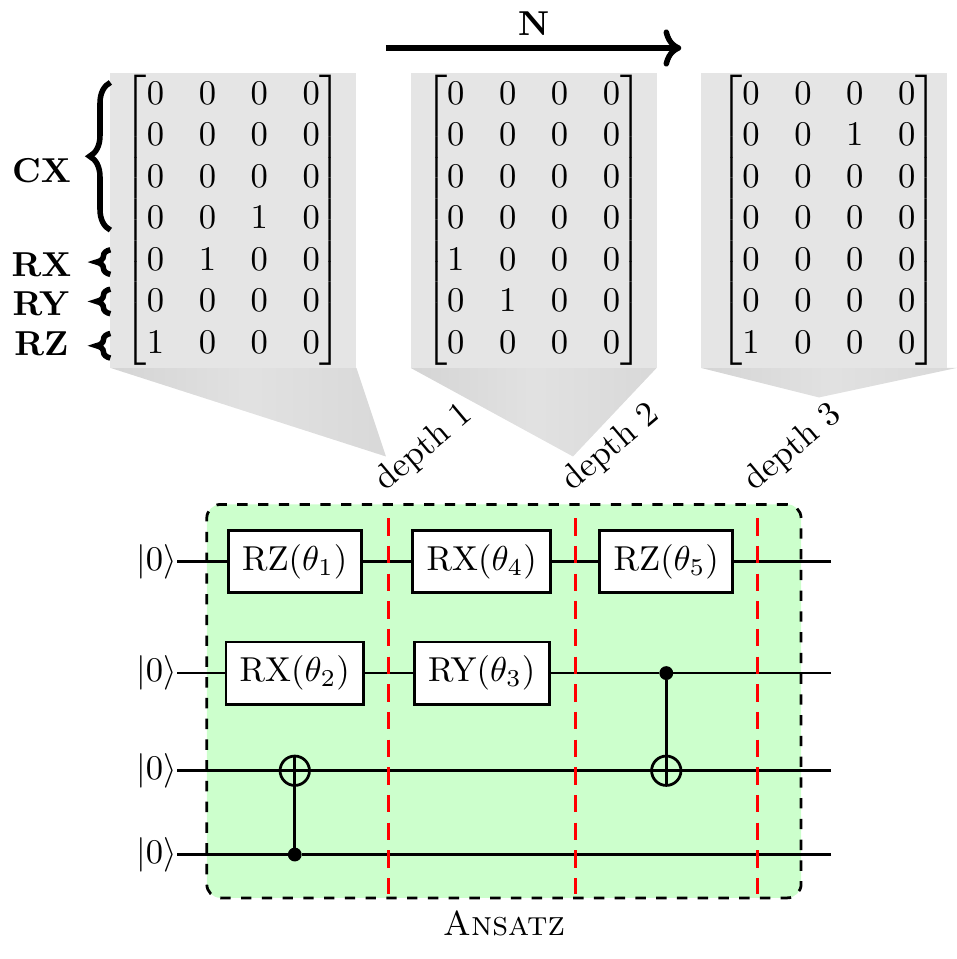}
\caption{\textbf{Example of the proposed encoding for a $4$-qubit ansatz into an RL-state}. The first $\left(N\times N\right)$ square matrix is reserved for the \texttt{CNOT} connectivity. The columns of the square matrix encode the \textit{target qubit}, and the rows represent \textit{control qubits}. The remaining $\left((N+j)\times N\right)$ elements encode arbitrary rotation towards $j$ direction where $j = 1$, $2$, and $3$, for $\texttt{X}$, $\texttt{Y}$ and $\texttt{Z}$ rotations, respectively.}
\label{fig:encoding}
\end{figure}
    
    \item \textbf{Depth-based encoding} In previous work~\cite{ostaszewski2021reinforcement} each $\left((N+3)\times N \right)$ matrix carries information corresponding to each action taken by the agent, where each action represents either a single or a two-qubit gate. Additionally, the information was integer-based, in the range $0$ to $N$.
    
    On the contrary, In our work, the encoding is binary and depth-based. For example, if $D_\textrm{max}=3$, then the encoding initiates by filling up the~$\left[i \times ((N + 3) \times N )\right]$ for $i =1$ until a depth of RL-ansatz is encoded. Thus, we have
    \begin{equation}
        \left[ \underbrace{ ((N + 3) \times N )}_{\textrm{depth = 1}}, \underbrace{((N + 3) \times N )}_{\textrm{all zeros}},\underbrace{((N + 3) \times N )}_{\textrm{all zeros}} \right].
    \end{equation}
    Then, as $i=1$ is filled up, we move to $i=2$ to encode depth = 2 of the RL-ansatz, which yields 
    \begin{equation}
        \left[ \underbrace{ ((N + 3) \times N )}_{\textrm{depth = 1}}, \underbrace{((N + 3) \times N )}_{\textrm{depth = 2}},\underbrace{((N + 3) \times N )}_{\textrm{all zeros}} \right].
    \end{equation}
Finally, the depth = 3 is encoded in $i=3$ resulting in
\begin{equation}
        \left[ \underbrace{ ((N + 3) \times N )}_{\textrm{depth = 1}}, \underbrace{((N + 3) \times N )}_{\textrm{depth = 2}},\underbrace{((N + 3) \times N )}_{\textrm{depth = 3}} \right].
    \end{equation}
\end{enumerate}
Each depth encoding follows the scheme shown in \figref{fig:encoding}.

\subsection{Actions}\label{sec:action_encoding}
For constructing the quantum circuits, we use the scheme developed in~\cite{ostaszewski2021reinforcement} with \texttt{CNOT} and one-qubit rotation gates, which are feasible on currently available quantum devices. The encoding of the action space can be defined as follows. The \texttt{CNOT} gates are represented by a pair of values that indicate the positions of the control and target qubits, with enumeration starting from 0. As for the rotation gates, they are encoded using two integers, also starting from 0. The first integer identifies the qubit register, while the second integer specifies the rotation axis. For an N-size quantum state,  the agent can choose from $3\times N$ single-qubit gates and $2\times\binom{N}{2}$ two-qubit gates. 
{As we are utilizing deep RL methods, we employ the one-hot encoding technique to represent actions within the action space. 
Mathematically, one-hot encoding can be identified as the Kronecker-Delta function as follows~\cite{guo2016entity}. Suppose $x$ is a discrete categorical random variable that takes $n$ distinct values $x_1, \ldots, x_n$. Then the one-hot encoding of a particular value $x_i$ is a vector $v$ where every component of $v$ is zero except for the $i$th component, which has the value $1$.
We refer the reader to Fig.~\ref{fig:encoding} for a visual illustration of one-hot encoding for the actions (shaded in grey color).
}

\subsection{Reward function}

To guide the agent quickly towards the goal, we introduce a reward that is dense in time at each time step $t$. The reward used in this work is given as
\begin{equation}
    R = \left\{\begin{array}{ll}
        +\mathcal{R} & \text{for } C_t(\vec{\theta})<\zeta+10^{-5}\\
        -\textrm{log}\left(C_t(\vec{\theta})-\zeta\right) & \text{for } C_t(\vec{\theta})>\zeta
        \end{array}\right.,\label{eq:log_reward}
\end{equation}
where the goal of the agent is to reach the minimum error for a predefined threshold $\zeta$, \ie the tolerance for cost function minimization. The $\zeta$ is a hyperparameter of the model. The cost function at each step $t$ is calculated for the ansatz which outputs a state $\rho_t(\vec{\theta})$ as
\begin{equation}
C_t(\vec{\theta}) = \text{Tr}(\rho^2) - \text{Tr}(\rho_t(\vec{\theta})^2).
\end{equation}

\subsection{Agent and environment specification}

In this work, we use a Double deep Q-network~\cite{mnih2013playing} (DDQN) for better stability with an $\epsilon$-greedy policy and the ADAM optimizer~\cite{kingma2014adam} {to optimize the weights of the neural network}. More details about the RL procedure are described in the next section. As mentioned in the previous section, to obtain a reward $R$ for the circuit (\ie for each environmental state), an optimization subroutine needs to be applied to determine the values of the rotation gate angles. We use well-developed methods for continuous optimization, such as Constrained Optimization By Linear Approximation~\cite{powell1994direct} (COBYLA), {which we utilize to optimize the parameters of the quantum circuit}. 

\section{Numerical demonstrations}\label{sec:results}

\paragraph{Setup details}
We start with the parameter specifications given in~\cite{ostaszewski2021reinforcement}, which uses the DDQN algorithm with a discount factor of $\gamma = 0.88$ and an $\epsilon$-greedy policy for selecting random actions. The value of $\epsilon$ is gradually decreased from $1$ to a minimum value of $0.05$ by a factor of $0.99995$ at each step. The size of the memory replay buffer is set to $2 \times 10^4$, and the target network in the DDQN training is updated with every $500$ action. Following each training episode, we conduct a testing phase where the probability of selecting a random action is set to $0$, and the experience replay procedure is turned off. Experiences obtained during the testing phase are not added to the memory replay buffer. The source code and the specifications of the numerical experiments presented in this section are available from the publicly accessible code repository~\cite{rl-vqsd-code}.

{\paragraph{Experiment details}
In the following numerical simulations, we consider $2$-qubit random quantum states and the reduced ground state of a $3$-qubit Heisenberg model to benchmark the performance of RL-VQSD in comparison with the VQSD method with $l$ layers of LHEA. Further, to show the scaling of RL-VQSD we consider diagonalizing the reduced ground state of the $4$-qubit Heisenberg model.
For the experiment, we consider $10000$ episodes (if not stated otherwise) where each episode is decomposed into $N_s$ steps. The value of $N_s$ is set to $20$, $40$ and $60$ while diagonalizing $2$, $3$ and $4$-qubit problems respectively. In each step of an episode, the RL-agent decides on an action following the encoding provided in section~\ref{sec:action_encoding}, and then the action is translated into either rotation or \texttt{CNOT} gate.
The value of the parameter for a new rotation gate is always initialized with $0$. Then, the parameters of the circuit are optimized. In the next step, when the RL-agent decides on a new action, the new rotation is initialized to $0$, but the previous rotation gates are set to their optimized angles. Then, the modified ansatz is optimized in the classical optimization subroutine (using COBYLA optimizer). This process is repeated until the problem is solved or all the steps in an episode are exhausted.
Mainly for this work, at each step of an episode, we optimize all angles at once (global strategy), which we call global COBYLA. In diagonalizing $2$-, $3$- and $4$-qubit quantum state we use $400$, $500$ and $1000$ iterations of COBYLA optimizer respectively.
}

{
\subsection{Analysis of RL-state encoding and reward function}
\begin{table}[h!]
\centering
\begin{tabular}{c|ccccc}
\hline
Settings & Qubits & Avg. err. & Avg. 1q gate & Avg. 2q gate & Avg. depth  \\
\hline
\multirow{2}{10em}{Binary encoding~\ref{sec:encoding} and dense reward~\ref{eq:log_reward}} & 2 & $5.23\times10^{-6}$ & $14.40$ & $4.60$ & $14.12$ \\
& 3 & $9.16\times10^{-5}$ & $18.19$ & $12.08$ & $20.32$ \\
\hline
\multirow{2}{10em}{Integer encoding and sparse reward~\cite{ostaszewski2021reinforcement}} &2 & $6.51\times10^{-6}$ & $13.96$ & $4.72$ & $14.81$  \\
& 3 & $5.97\times10^{-4}$ & $15.49$ & $24.50$ & $35.65$ \\
\hline
\end{tabular}
\caption{{\textbf{Comparison of the binary encoding along with the reward function presented in the paper with the integer-based encoding and the sparse reward function presented in~\cite{ostaszewski2021reinforcement}}. We diagonalize a $2$-qubit arbitrary state and the reduced ground state of the $3$-qubit Heisenberg model. The result shows that for $2$-qubit, both the settings perform equivalently, but as we scale up the system to $3$-qubit, the integer encoding with sparse reward fails to give an efficient RL-ansatz with small gates and depth that can help us achieve a good approximation of the eigenvalues. This study leads us to conclude that the success of an RL-agent significantly depends on appropriately encoding the RL-state and designing the reward function.
}}
\label{tab:settings_comparison}
\end{table}
Before diving into a rigorous investigation of the performance of RL-VQSD, we showcase the effectiveness of the RL-state encoding method (provided in section~\ref{sec:encoding}) along with the dense reward function (as in Eq.~\ref{eq:log_reward}). To benchmark the effectiveness, we
compare the RL-ansatz proposed by the agent utilizing the following two settings to diagonalize $2$- and $3$-qubit quantum state with RL-VQSD: (1) the binary encoding scheme along with the dense reward function presented in this paper and (2) the integer-based RL-state encoding scheme (which we call \textit{integer encoding}) and the sparse reward function described in~\cite{ostaszewski2021reinforcement} in solving quantum chemistry problems. In both cases, we keep the agent and environment specifications unchanged.
Through this investigation, we show how the RL-state encoding and the engineering of the reward function are responsible for the success of the RL-VQSD method.
In Table~\ref{tab:settings_comparison}, we present the results, which confirm that the integer encoding, along with sparse reward in~\cite{ostaszewski2021reinforcement}, underperforms in finding a more accurate diagonalization of the state with a smaller number of gates and depth as the size of the diagonalizing state increases. Furthermore, it does not solve the diagonalization problem with a $10^{-4}$ threshold, a problem easily tackled by binary encoding and the dense reward methods.}

\subsection{$2$-qubit random quantum states}
\begin{figure}[h!]
\centering
\begin{subfigure}{.5\textwidth}
  \centering
  \includegraphics[width=\linewidth]{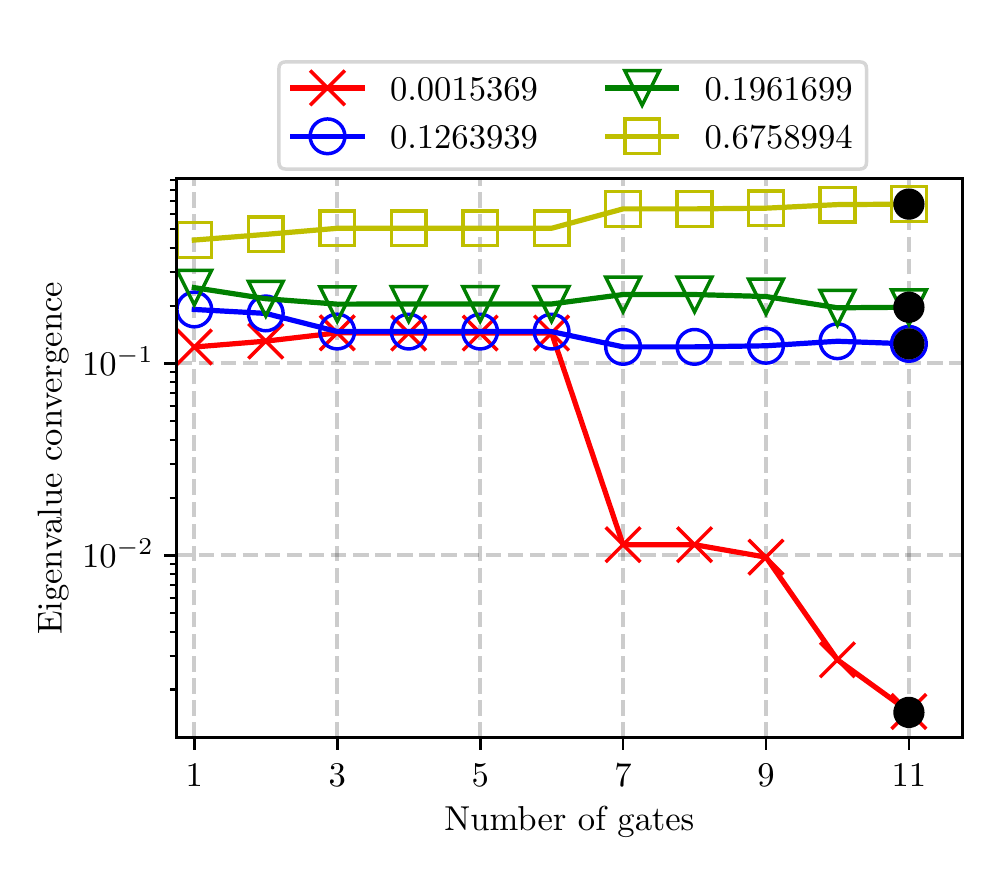}
  \caption{Single mixed quantum state.}
  \label{fig:2-qubit-eigenvalue-convergence}
\end{subfigure}%
\begin{subfigure}{.5\textwidth}
  \centering
  \includegraphics[width=\linewidth]{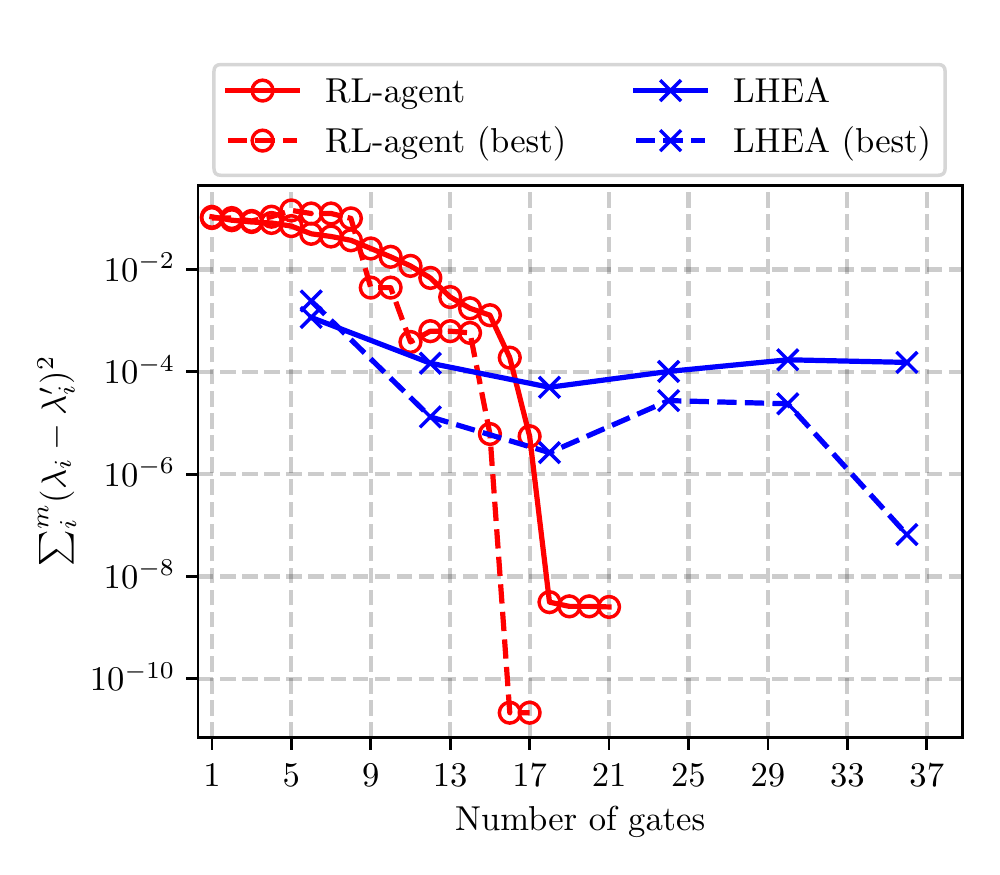}
  \caption{Average over $50$ random quantum state.}
  \label{fig:2_qubit_average_error_50_states}
\end{subfigure}
\caption{\textbf{The summary of results for diagonalizing full rank $2$-qubit random density matrix}. In (\subref{fig:2-qubit-eigenvalue-convergence}) we illustrate eigenvalue convergence for the diagonalization of a single mixed quantum state. In  (\subref{fig:2_qubit_average_error_50_states}) we compare the performance of the RL-agent-generated ansatz with the LHEA. It can be seen that the RL-agent-generated ansatz gives us a better approximation of the eigenvalues. Additionally, the RL-based methods can achieve the accuracy of the LHEA using the circuit with significantly reduced depth of the resulting circuit.}
\label{fig:2_qubit_eig_conv_HEA_comparison}
\end{figure}

In the first numerical experiment, we utilize RL-VQSD to diagonalize (1) a single mixed quantum state and (2) $50$ random quantum states of the full rank of $2$-qubit, to get the average eigenvalue approximation error and count the gates in RL-ansatz. We utilized the \texttt{random\_density\_matrix} of the module \texttt{quantum\_info} of \texttt{qiskit}~\cite{qiskit2019} to sample the quantum states from the Haar measure. By (1), we argue that RL-VQSD can exactly diagonalize a quantum state. The results of (2) demonstrate that the average performance of RL-VQSD is better than state-of-the-art ansatz.
\begin{figure}[t!]
    \centering	\includegraphics{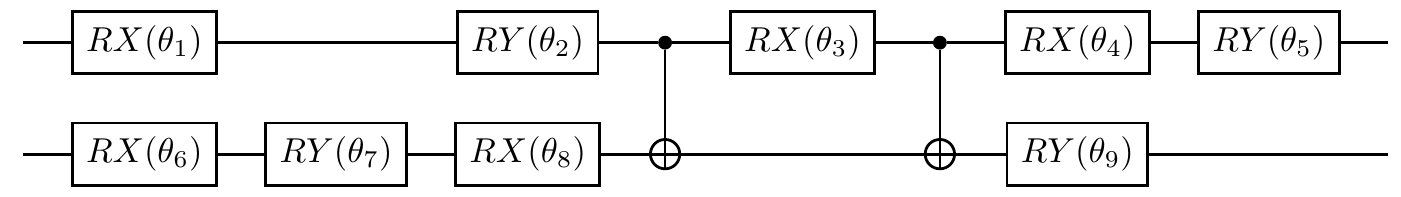}
	\caption{\textbf{The ansatz proposed by RL-agent to diagonalize the $2$-qubit state with eigenvalues convergence illustrated in \figref{fig:2-qubit-eigenvalue-convergence}}. This shows us that even with very few gates and small depth the RL-VQSD can give us accurate diagonalization of small quantum systems.}
    \label{fig:2_qubit_fixed_depth_ansatz}
\end{figure}

In \figref{fig:2-qubit-eigenvalue-convergence} we show that the agent can propose an ansatz that provides us with the exact eigenvalues for a $2$-qubit random quantum state with $12$ gates, containing $10$ rotations and $2$ \texttt{CNOT} gates. The RL-ansatz is depicted in \figref{fig:2_qubit_fixed_depth_ansatz}.
\begin{figure}[h!]
\centering
\includegraphics[width=0.65\textwidth]{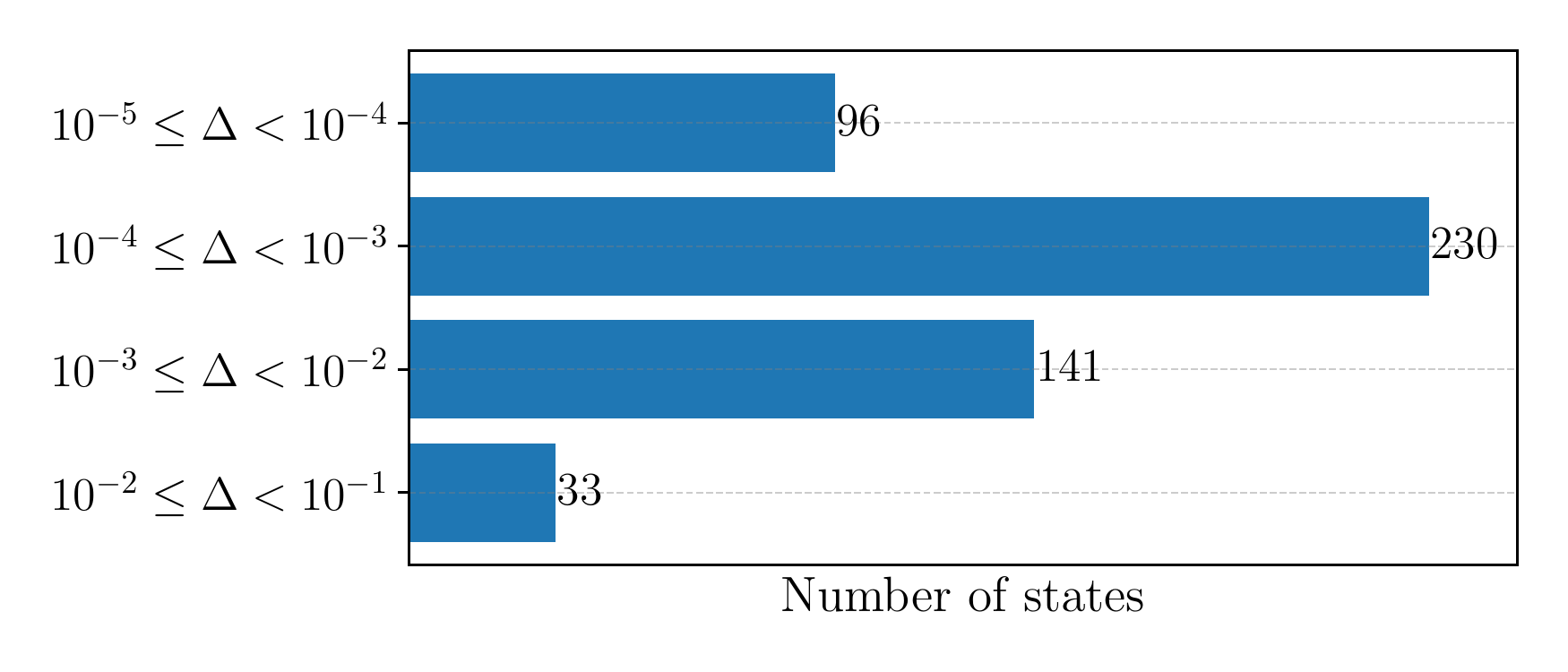}
\caption{\textbf{Statistics of error in eigenvalue estimation for $500$ arbitrary quantum states}. As an ansatz to diagonalize all the $500$ random quantum states, we consider a fixed structure provided by the RL-agent. In this case, we consider the structure given in \figref{fig:2_qubit_fixed_depth_ansatz}.}
\label{fig:2_qubit_error_with_fixed_ansatz}
\end{figure}

Meanwhile, in \figref{fig:2_qubit_average_error_50_states}, we benchmark the performance of RL-ansatz against LHEA. In the illustration, we show that the agent not only gives us a small ansatz to diagonalize with a specific predefined threshold $\zeta=10^{-5}$ but also helps us achieve a lower error in eigenvalue estimation compared to LHEA.
{Furthermore, in Table~\ref{tab:2q-VQSD_and_RL_VQSD_comparison}, we provide a rigorous comparison of the RL-ansatz and $6$ layers of LHEA (of the structure depicted in Fig.~\ref{fig:layered-anzatz-three}).
For the comparison, we evaluate the average statistical error (which is defined in Eq.~\ref{eq:statistical_error}) in estimating eigenvalues, the mean count of one and two-qubit gates, the average depth, and the overall average count of gates as metrics.
From the table, we can conclude that the average error of LHEA gets stuck around $10^{-5}$, where the 2nd and the 3rd layers of LHEA provide the lowest error in eigenvalue estimation. Meanwhile, the RL-ansatz can, on average, give $10^{2}$ times less error compared to LHEA with an ansatz composed of $3$ times fewer parameterized gates and smaller depth. In Table~\ref{tab:2q-VQSD_and_RL_VQSD_comparison}, for depth $14$ LHEA (which corresponds to the 2nd layer of LHEA with $26$ gates), we see that the average error in eigenvalue reaches to $1.31\times10^{-5}$. On the other hand, for the same depth, the RL-ansatz can achieve an error of $9.33\times10^{-7}$ with an average (over the $50$ random states) of $2.56$ $2$-qubit and $11.58$ $1$-qubit gates. }

Furthermore, we explore the possibility of utilizing the ansatz proposed by the RL-agent, trained on a specific quantum state, to diagonalize random quantum states that differ from the initial fixed state. We can confirm that this is indeed possible in the case of the $2$-qubit state. The corresponding results are presented in \figref{fig:2_qubit_error_with_fixed_ansatz}. One can argue that, in this case, the diagonalization task is relatively easy. However, our results show that it is possible to harness the RL-ansatz for a particular quantum state to diagonalize an arbitrary state of the same dimension. To conduct this experiment, we start by selecting a random quantum state and training it using the RL-agent. The RL-agent then provides us with an RL-ansatz specifically designed for that particular state. By utilizing this RL-ansatz as a quantum circuit and employing VQSD (refer to \figref{fig:ill-vqsd}), we successfully diagonalize $500$ arbitrary quantum states. Our results indicate that the RL-ansatz achieves a reasonable accuracy, with the majority of quantum states falling within the range of $10^{-4}\leq \Delta \leq 10^{-3}$.
\begin{table}[h!]
\centering
\begin{tabular}{c|cccccc}
\hline
 & Layer & Avg. error & Avg. 1q gate & Avg. 2q gate & Avg. depth & Avg. total gate \\ \hline
\begin{tabular}[c]{@{}l@{}}
RL-ansatz
\end{tabular}
& N.A. & $9.33\times10^{-7}$ & $10.68$ & $2.28$ & $9.54$ & $14.06$ \\
\hline
\multirow{6}{4em}{6 layers of LHEA}          
& 1 & $2.42\times10^{-4}$ & $12$ & $1$ & $7$  & $13$\\
& 2 & $1.31\times10^{-5}$ & $24$ & $2$ & $14$ & $26$\\
& 3 & $4.98\times10^{-5}$ & $36$ & $3$ & $21$ & $39$\\
& 4 & $1.02\times10^{-4}$ & $48$ & $4$ & $28$ & $52$\\
& 5 & $1.72\times10^{-4}$ & $60$ & $5$ & $35$ & $65$\\
& 6 & $1.53\times10^{-4}$ & $72$ & $6$ & $42$ & $78$\\
\hline
\end{tabular}
\caption{{\textbf{Comparison of the RL-ansatz (proposed by the RL-agent) and $6$ layers of LHEA (utilized in the VQSD) used for diagonalizing $2$-qubit arbitrary quantum states}. For the comparison, we consider investigating the average error in eigenvalue estimation (Avg. error), the average number of single-qubit gates (Avg. 1q gate), the average number of the two-qubit gate (Avg. 2q gate), the average depth (Avg. depth) and the average number of total gates (Avg. total gate). The structure of the LHEA (depicted in Fig.~\ref{fig:layered-anzatz-three}) utilized for the investigation is of a fixed structure whose depth and gates scales as $l\times (\text{depth or number of gates})$ where $l$ denotes the layers of the LHEA. It should be noted that the average is taken over $50$ random quantum states. N.A. denotes not applicable.}}
\label{tab:2q-VQSD_and_RL_VQSD_comparison}
\end{table}

\subsection{$3$-qubit reduced Heisenberg model}
One of the important applications of VQSD is to study the entanglement in condensed matter systems~\cite{li2008entanglement}. Hence, in this experiment, to get a better understanding of the efficacy of our method in this regard, we consider a $3$-qubit reduced state of the ground state ($|\psi_{S_1,S_2}\rangle$) of the one-dimensional Heisenberg model defined on six qubits which have the following form
\begin{equation}
	H=\sum_{j=1}^{2N} \vec{S}^{(j)}\cdot\vec{S}^{(j+1)},
\end{equation}
where $\vec{S}^{(j)} = \frac{1}{\sqrt{3}}\left( X^{(j)} \hat{x}+Y^{(j)} \hat{y}+Z^{(j)} \hat{z} \right)$ with periodic boundary condition $\vec{S}^{(2N+1)} = \vec{S}^{(1)}$, where $X$, $Y$, and $Z$ are the Pauli operators. To perform entanglement spectroscopy on the ground state of the 6-spin Heisenberg model (\ie $2N=6$), we diagonalize the reduced state $\rho_\textrm{red} = \textrm{Tr}_{S_2}\left[ |\psi_{S_1,S_2}\rangle\langle\psi_{S_1,S_2}| \right]$. We set the predefined threshold $\zeta = 10^{-4}$. {We decided to choose a higher value of $\zeta$ compared to the value considered for the $2$-qubit problem because as we increase the number of qubits, the problem of diagonalizing quantum states becomes more difficult, leading to complicated structures of RL-ansatz. Hence, we can choose a higher value of $\zeta$ to lower the difficulty. In Appendix~\ref{appndix:threshold_dependence}, we elaborate on how the number of gates and the depth of the RL-ansatz varies as we make the problem more difficult by lowering the $\zeta$.}

\begin{figure}[h!]
    \centering \includegraphics[width = 0.5\textwidth]{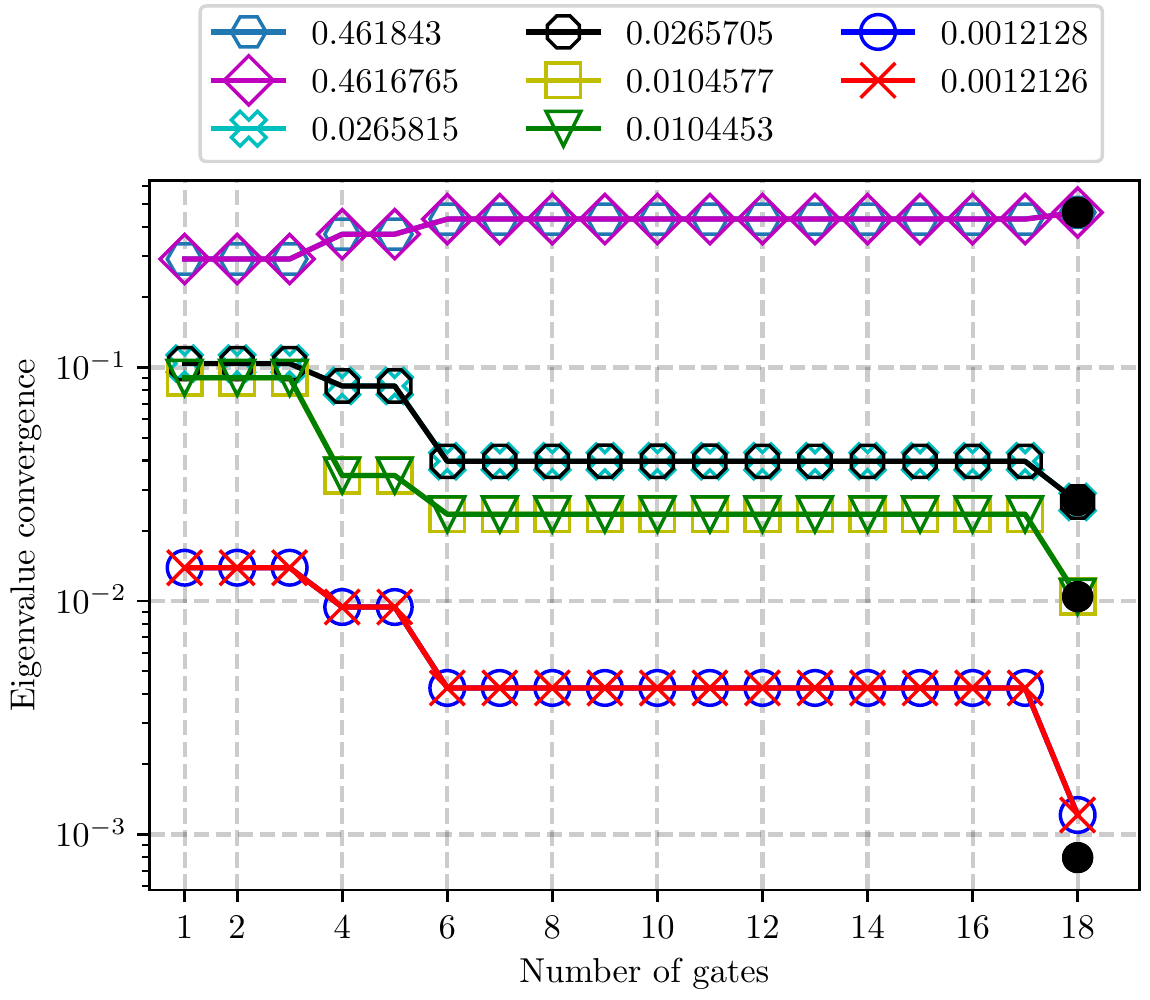}
    \caption{\textbf{Convergence of the eigenvalues of the reduced ground state of $3$-qubit Heisenberg model by RL-VQSD}. The labels on the top of the figure correspond to the different eigenvalues. The black dots in the plot represent the true eigenvalues. There are eight black dots. However, as some of the eigenvalues coincide up to three decimal places, they are indistinguishable.}
    \label{fig:3_qubit_reduced_heisenberg_model}
\end{figure}

The results presented in \figref{fig:3_qubit_reduced_heisenberg_model} confirm that the RL-agent can learn to construct an ansatz to find all the eigenvalues with reasonable accuracy. In this case, one can see that the ansatz takes 18 quantum gates to give us 6 out of 8 exact eigenvalues of a $3$-qubit Heisenberg model. Additionally, the RL-ansatz finds the remaining two smallest eigenvalues with $1.73\times 10^{-7}$ accuracy. In \figref{fig:3_qubit_reduced_heisenberg_model_ansatz}, we present the RL-ansatz that contains $10$ rotations and $8$ \texttt{CNOT} gates proposed by the RL-VQSD.
{In the Table~\ref{tab:3q-VQSD_and_RL_VQSD_comparison}, we investigate the performance of RL-ansatz (proposed by the RL-agent) and $4$ layers of LHEA (used in VQSD) to solve $3$-qubit Heisenberg model. As the metrics for the comparison, we evaluate the minimum statistical error in estimating eigenvalues, the minimum count of one and two-qubit gates, the minimum depth, and the overall minimum count of gates. It can be seen that the RL-ansatz can give us $10$ times lower energy compared to LHEA with $4$ layers. Meanwhile, the RL-ansatz comprises more than $3$ times fewer parameters to achieve this accuracy. This clearly shows that the RL-ansatz is more efficient than the LHEA in the VQSD task and returns a smaller error in eigenvalue estimation. We also see that for depth $21$ LHEA (which corresponds to the layer $1$ with $39$ gates), the average error in eigenvalue reaches $4.59\times10^{-4}$. On the other hand, for the same depth, the RL-ansatz can achieve an error of $2.43\times10^{-5}$ with an average (over all the successful episodes) of $8$ two-qubit gates and $14$ one-qubit gates.}

\begin{figure}[tbh!]
    \centering	\includegraphics[width = \textwidth]{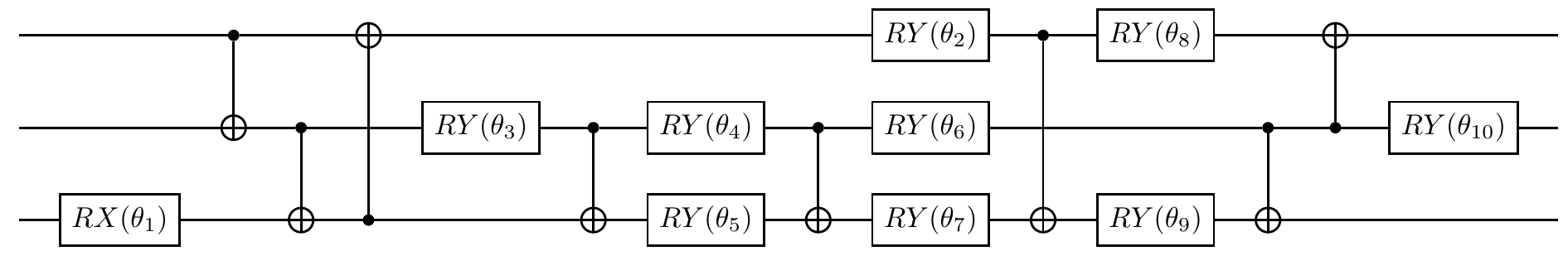}
    \caption{\textbf{The ansatz proposed by the RL-agent for diagonalizing a state in the $3$-qubit reduced Heisenberg model}. The circuit contains $10$ rotations and $8$ \texttt{CNOT} gates.}
\label{fig:3_qubit_reduced_heisenberg_model_ansatz}
\end{figure}
\begin{table}[]
\centering
\begin{tabular}{c|cccccc}
\hline
 & Layer & Min. error & Min. 1q gate & Min. 2q gate & Min. depth & Min. total gate \\ \hline
\begin{tabular}[c]{@{}l@{}}
RL-ansatz
\end{tabular}
& N.A. & $2.43\times10^{-5}$ & $10$ & $8$ & $12$ & $18$ \\
\hline
\multirow{4}{4em}{$4$ layers of LHEA}          
& 1 & $4.59\times10^{-4}$ & $36$ & $3$ & $21$  & $39$\\
& 2 & $3.63\times10^{-4}$ & $72$ & $6$ & $42$ & $78$\\
& 3 & $4.35\times10^{-4}$ & $108$ & $9$ & $63$ & $117$\\
& 4 & $5.82\times10^{-4}$ & $144$ & $12$ & $84$ & $156$\\
\hline
\end{tabular}
\caption{{\textbf{Comparison of the RL-ansatz (proposed by RL-agent) with $4$ layers of LHEA (utilized in VQSD) used for diagonalizing $3$-qubit Heisenberg model}. For the comparison, we consider investigating the minimum error in eigenvalue estimation (Min. error), the minimum number of single-qubit gates (Min. 1q gate), the minimum number of the two-qubit gate (Min. 2q gate), the minimum depth (Min. depth) of the ansatz and the minimum number of total gates (Min. total gate). N.A. denotes not applicable.}}
\label{tab:3q-VQSD_and_RL_VQSD_comparison}
\end{table}

It should be noted from circuits in \figref{fig:2_qubit_fixed_depth_ansatz} and in \figref{fig:3_qubit_reduced_heisenberg_model_ansatz} that the rotation in the \texttt{Z} direction, \ie \texttt{RZ} quantum logic gate, does not play a crucial part in the diagonalizing unitary. Thus, one might attempt to diagonalize a random quantum state of two and three qubits, excluding \texttt{RZ} rotation from the list of quantum gates. This gives us a hint concerning the action space that could be significantly reduced in these examples.

\subsection{$4$-qubit reduced Heisenberg model}
We extend the results of the previous section for the ground state of $8$-spin Heisenberg model (\ie $2n = 8$). We diagonalize the $4$-qubit reduced state of the ground state of the $8$-spin Heisenberg model. 
\begin{figure}[h!]
    \centering
    \includegraphics[width=\textwidth]{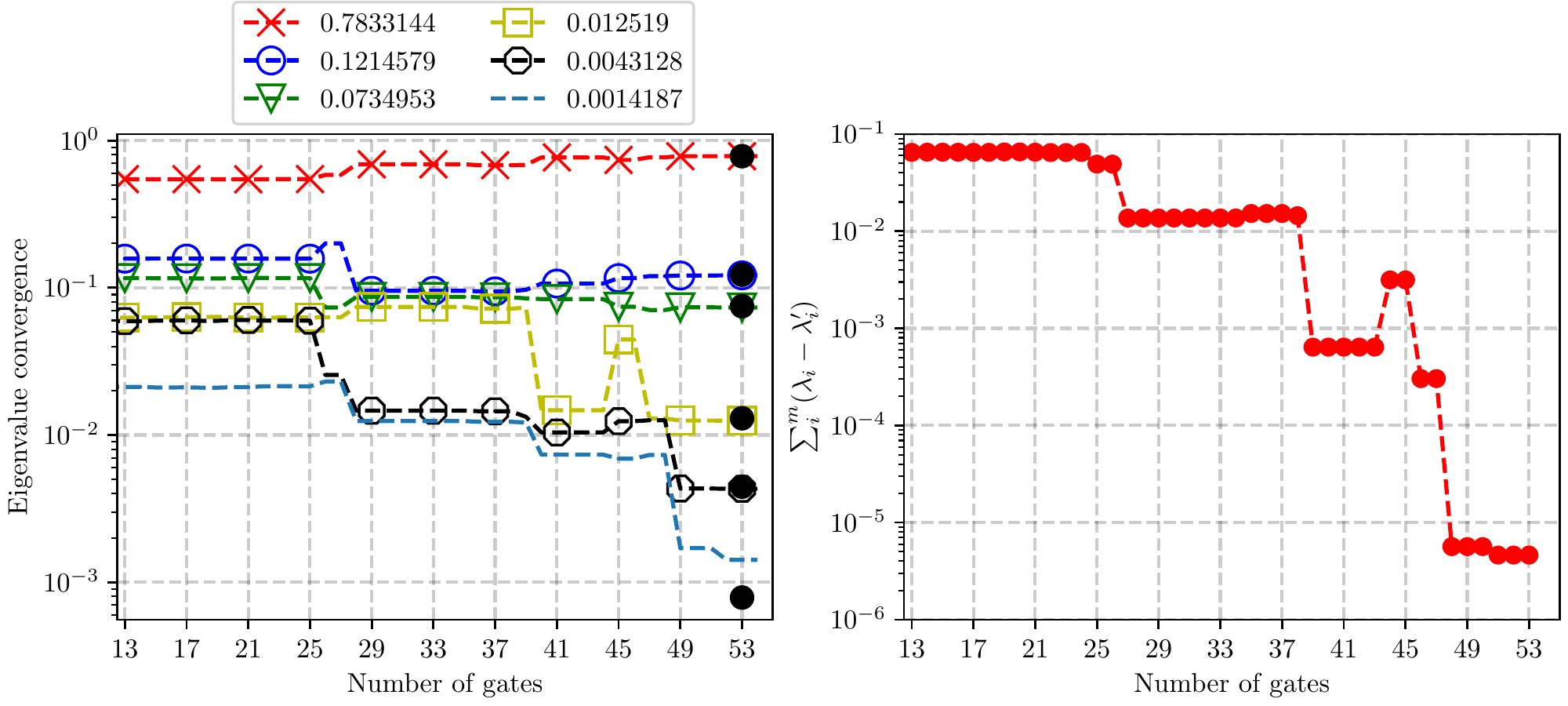}
    \caption{\textbf{The convergence of individual (left panel) and the overall error (right panel) in the estimation of eigenvalues for the reduced ground state of $4$-qubit Heisenberg model}. This provides a significant improvement in terms of gate count and depth compared to the result reported in~\cite{larose2019variational}.}
    \label{fig:4-qubit_rl_vqsd_convergence}
\end{figure}

The convergence of the eigenvalues is illustrated in the \figref{fig:4-qubit_rl_vqsd_convergence}. For our investigation, we show that it takes $53$ gates to find the first $6$ largest eigenvalues with an error below $10^{-5}$. Out of $53$ gates, $16$ are \texttt{CNOT}, and the remaining are $1$-qubit rotations. Throughout the experiment, we consider choosing the predefined threshold $\zeta = 10^{-3}$.
\begin{figure}[h!]
\begin{subfigure}[b]{\textwidth}
	\centering
	\includegraphics{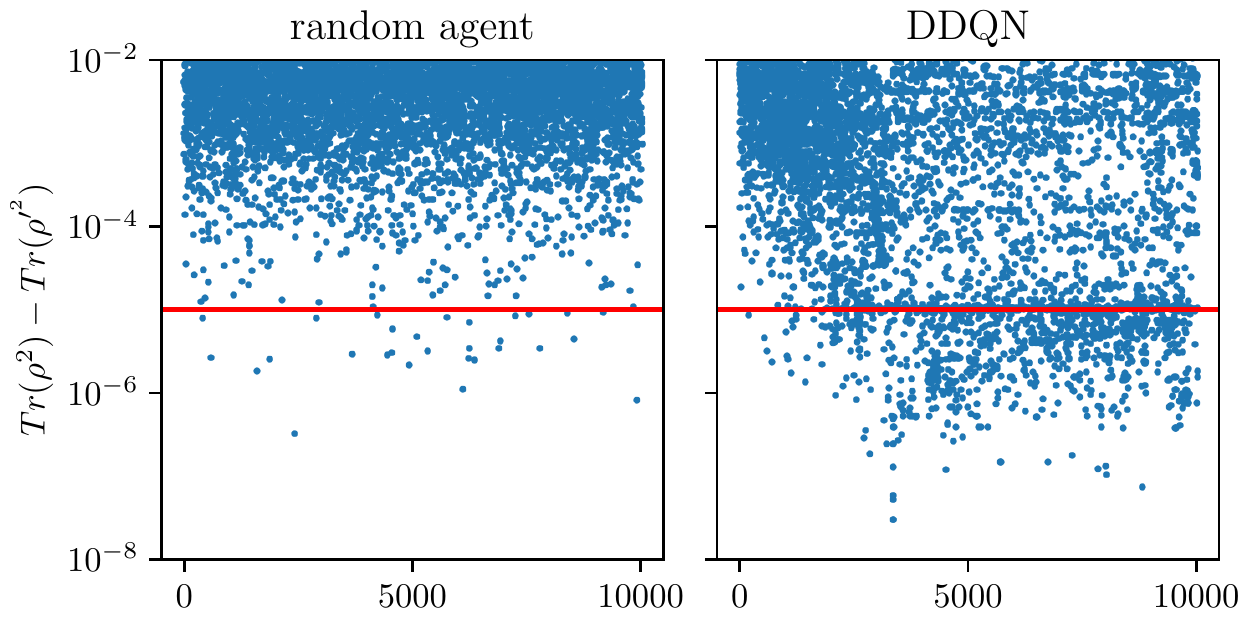}
	\caption{}
    \label{fig:random_search_vs_agent_2_qubit}
\end{subfigure}
\begin{subfigure}[b]{\textwidth}
\centering	\includegraphics{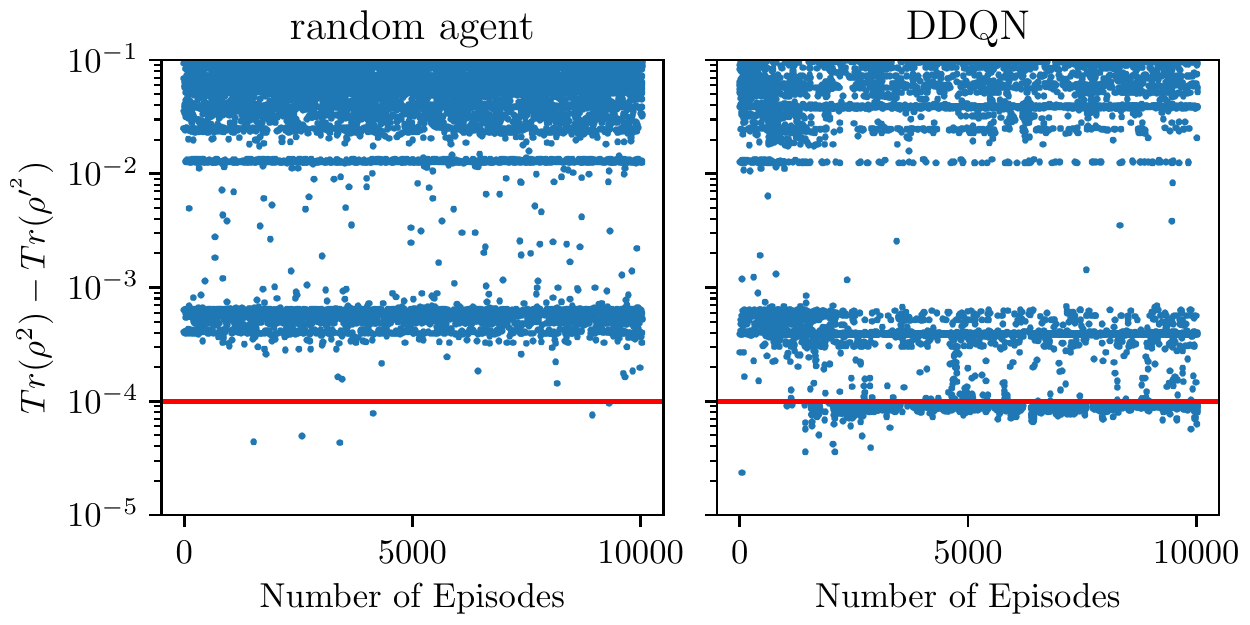}
	\caption{}
    \label{fig:random_search_vs_agent_3_qubit}
    \end{subfigure}
\caption{\textbf{Comparison of accuracy obtained using a random agent and RL-based method in diagonalizing $2$ and $3$-qubit states}. For this simulation, we utilized $10^4$ episodes to solve the full rank (\subref{fig:random_search_vs_agent_2_qubit}) random quantum state of $2$-qubit and (\subref{fig:random_search_vs_agent_3_qubit}) the reduced ground state of $3$-qubit Heisenberg model. It can be seen that the RL-agent can give us more frequent solutions, whereas the random agent can hardly solve the problem. The red line denotes the predefined tolerance for the approximation of the cost function.}
\label{fig:random_search_vs_agent}
\end{figure}

\begin{table}[h!]
\centering
\begin{tabular}{c|ccc}
\hline
Qubits & Min. depth & Min. 1q gate & Min. 2q gate \\
\hline
2 & 8  & 9 & 2              \\
3 & 12 & 10 & 8              \\
4 & 33 & 28 & 16\\
\hline
\end{tabular}
\caption{\textbf{Summary of the required minimum number of one (min. 1q gate), the minimum number of two-qubit gates (min. 2q gate) required, and the depth (min. depth) in RL-ansatz to diagonalize $2$-, $3$- and $4$-qubit systems}. To gather this data we run $10^4$ episodes of the RL-VQSD for each qubit case utilizing the settings provided in setup and experimental details in the first two paragraphs of section~\ref{sec:results}.}
\label{tab:resource-count}
\end{table}

The summary of our results is provided in Table~\ref{tab:resource-count}. One can notice that there is a relation between the number of \texttt{CNOT}s and the dimension of the state that we want to diagonalize. The number of \texttt{CNOT}s grows exponentially with the number of qubits. As for the two-qubit case, we find all the eigenvalues with $10^{-10}$ error with just two \texttt{CNOT}s. Whereas for three qubits, we can find the first $6$ eigenvalues with an error below $10^{-8}$ but the smallest two eigenvalues we find with $1.73\times10^{-7}$ error with $8$ \texttt{CNOT}s. Finally, for $4$-qubit, we see the first $6$ eigenvalues with an error below $10^{-8}$ and the remaining eigenvalues with an error in the range $10^{-4}\leq\Delta\leq 10^{-6} $ with $16$ \texttt{CNOT}s. This observation suggests that for a full-rank quantum state of $N\geq 3$, we require at least as many \texttt{CNOT}s as the rank of the quantum state to get a good approximation of the largest eigenvalues. It should be noted that to find the first $5$ largest eigenvalues with error $10^{-5}$ the ansatz proposed by the RL-agent is of depth 18 and a total of $30$ gates, among which 12 are \texttt{CNOT} gates and the remaining are rotations. This significantly improves the depth, and the gate count in the diagonalizing ansatz compared to the results in~\cite{cerezo2022variational} and \cite{larose2019variational}.

\subsection{Performance of random agent}
To demonstrate the hardness of the variational diagonalization task, we utilize a random agent to find an efficient ansatz in this section. Unlike the previous examples where an RL-agent selects an action based on a policy, here in the random agent settings, the action at each step is chosen randomly from a uniform distribution.
\begin{figure}[h!]
\centering	
\includegraphics[width=0.5\textwidth]{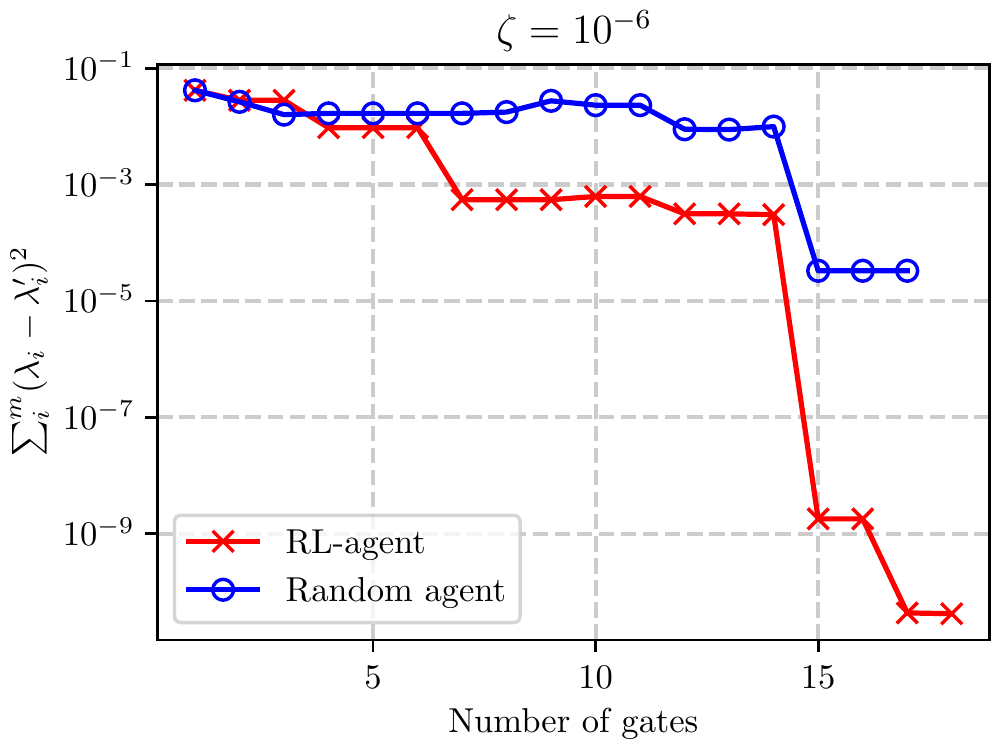}
\caption{\textbf{The comparison reaching accuracy in the order of $10^{-6}$ while diagonalizing $2$-qubit state by RL-agent and the random agent as a function of number of gates in the circuit}. To get the result we illustrate the variation of error in eigenvalue estimation with respect to the number of gates. It can be seen that the random agent halts after a certain error in eigenvalue estimation, whereas the RL-agent can go below the $10^{-6}$ in fewer gates.}
\label{fig:eigen_err_random_vs_rl_search}
\end{figure} 

In \figref{fig:random_search_vs_agent} (in the first column), we show the results for a random agent to diagonalize a $2$- and $3$-qubit quantum state. It can be seen that the number of successful episodes (the episodes that pass the predefined tolerance of cost function) drastically reduces as we scale the number of qubits in the state. At the same time, the RL-agent (in the second column) provides us with a more consistent outcome. {This occurs because in the scenario of a random agent, even though the RL process is active, the neural networks do not utilize the information (the reward) from the environment to determine the subsequent actions. Whereas in the case of the RL-agent each subsequent action is decided based on the cumulative reward received from the environment after each step. In Appendix~\ref{appndix:training_time} we investigate the training time for the RL-agent and show that the time it takes to complete an equal number of episodes by a random agent and an RL-agent is comparable.}
    
Additionally, from the results presented in \figref{fig:eigen_err_random_vs_rl_search}, one can conclude that even in the successful episodes, the number of gates in the ansatz proposed by the random agent is longer compared to RL-ansatz.
Hence, one can argue that a random agent cannot be reliably utilized to find an efficient ansatz for the VQSD, and a higher level of sophistication in learning is necessary to attain consistent results.

\section{Final remarks}\label{sec:final}
This paper proposed a novel method to construct the ansatz for the Variational Quantum State Diagonalization (VQSD) based on Reinforcement Learning (RL) and compared its performance with the conventional fixed-depth ansatz. To this end, we introduced an RL-based algorithm {that utilizes a novel binary encoding scheme and a dense reward function with the particular problem in mind. We showed that in solving the diagonalization problem the combination of the binary encoding and the dense reward function outperforms the previously proposed encoding and rewards proposed for solving quantum chemistry problems~\cite{ostaszewski2021reinforcement}. Indicating that proper engineering of the reward function and an efficient RL-state encoding is responsible for the agent's success.}
In particular, we show, for the VQSD task, a Double deep Q-network (DDQN) algorithm {with $\epsilon$-greedy policy} can be utilized to construct an ansatz (which is termed \textit{RL-ansatz}), shorter than the standard Linear Hardware Efficient Ansatz (LHEA). As such, compared to LHEA, the RL-ansatz is of smaller depth and a smaller gate count with better accuracy in the eigenvalue estimation. This makes RL-ansatz more suitable for implementation in near-term quantum devices. Hence, the provided numerical results suggest that our approach is suitable for improving the readiness of quantum computers in tasks related to quantum data processing. The proposed state encoding method and the reward function can be readily adapted to address various variational quantum algorithms. {It should be emphasized that like to emphasize that the RL-VQSD does not depend on the system size, and in principle can be used to diagonalize larger systems. However, we are currently limited by classical simulation capabilities and existing quantum devices. This is because the VQSD requires $2n$ number of qubits for $n$ size quantum state and the training time required to get the optimal ansatz increases rapidly with system size.}

Additionally, we demonstrated the hardness of the diagonalization task by replacing the RL-agent with a random agent, where the actions are chosen randomly from a uniform distribution. The results indicate that we can not reliably utilize a random agent in the diagonalization task as the number of successful episodes, where the cost function passes a predefined threshold, reduces rapidly as we scale up the size of the quantum state. Moreover, in the successful episodes, the random agent produces lengthy circuits compared to RL-agent. 

To summarize our contribution, we opened up the possibility of utilizing RL to explore the quantum state diagonalization problem. Compared to the previous works on VQSD, we show that RL can boost the performance of this procedure by reducing the number of gates in the diagonalizing ansatz. As such, it provides a viable method for increasing the readiness of the VQSD algorithm for implementation on near-term quantum computers. The possibility of harnessing the cost function landscape using other search algorithms remains an open problem.

\paragraph{Acknowledgements}
AK would like to thank Aritra Sarkar for the fruitful discussions.
AK and JM would like to acknowledge support from the Polish National Science Center under the grant agreement 2019/33/B/ST6/02011, and MO would like to acknowledge support from the Polish National Science Center under the grant agreement 2020/39/B/ST6/01511 and from Warsaw University of Technology within the Excellence Initiative: Research University (IDUB) programme. This research was partially supported by PL-Grid Infrastructure grant nr PLG/2023/016205. VD was supported by the Dutch Research Council (NWO/OCW), as part of the Quantum Software Consortium programme (project number 024.003.037). This work was also supported by the Dutch National Growth Fund (NGF), as part of the Quantum Delta NL programme. JM would like to thank Izabela Miszczak for proofreading the manuscript.

\paragraph{Competing Interest}
The authors have disclosed that they do not have any competing financial or non-financial interests.

\paragraph{Data Availability}
The code used to generate the results in the article is available at \cite{rl-vqsd-code}.

\paragraph{Author Contributions} AK, OD, PB, and YJP implemented the accompanying code. AK, PB, and JAM did the analysis. MB, JAM, and VD are responsible for the supervision, methodology, visualization, and preparation of the final manuscript. AK is responsible for the conceptualization.

\bibliographystyle{unsrturl}
\bibliography{rl_variational_diag}

\appendix

\section{Double deep Q-network}\label{app:ddqn}

{Deep RL methods employ neural networks to adapt the agent's policy for optimizing the return
$$
G_t=\sum_{k=0}^{\infty} \gamma^{k} r_{t + k + 1},
$$
with the discount factor $\gamma \in[0,1)$.
Each state and action pair $(s, a)$ can then be assigned an action-value that quantifies the expected return from state $s$ in step $t$ taking action $a$ under policy $\pi$
$$
q_\pi(s, a)=\mathbb{E}_\pi\left[G_t \mid s_t=s, a_t=a\right].
$$
The aim is to find the optimal policy that maximizes the expected return. 
Such a policy can be derived from the optimal action-value function $q_*$, defined by the Bellman optimality equation:
$$
q_*(s, a)=\mathbb{E}\left[r_{t+1}+\max_{a^{\prime}} q_*\left(s_{t+1}, a^{\prime}\right) \mid s_t=s, a_t=a\right].
$$
Instead of directly solving the Bellman optimality equation in value-based RL, the aim is to learn the optimal action-value function from data samples. 
One such prominent value-based $\mathrm{RL}$ algorithms is $Q$-learning, where each state-action pair $(s, a)$ is assigned a so-called $Q$-value $Q(s, a)$ which is updated to approximate $q_*$.
Starting from randomly initialized values, the Q-values are updated according to the following rule:
$$
Q(s_t, a_t) \leftarrow Q(s_t, a_t) + \alpha \left(r_{t+1} +\gamma \max_{a^{\prime}} Q\left(s_{t+1}, a^{\prime}\right) - Q(s_t, a_t)\right),
$$
where $\alpha$ is the learning rate, $r_{t+1}$ is the reward at time $t+1$, and $s_{t+1}$ is the next encountered state after taking action $a_t$ in state $s_t$. 
In the limit of visiting all $(s, a)$ pairs infinitely often, this update rule is proven to converge to the optimal Q-values in the tabular case~\cite{melo2001convergence}.
In practice, to ensure sufficient exploration in Q-learning setting, a so-called $\epsilon$-greedy policy is used. Formally, stated as,
$$
\pi(a \mid s)=\left\{\begin{array}{l}
1-\epsilon_t \text { for } a=\max _{a^{\prime}} Q\left(s, a^{\prime}\right) \\
\epsilon_t \text { otherwise }
\end{array}\right.
$$
The $\epsilon$-greedy policy is only used to introduce randomness to the actions selected by the agent during training, but once training is finished, a deterministic policy follows.}

{We employ neural networks (NN) as function approximators to extend Q-learning to large state and action spaces.
NN training typically requires independently and identically distributed data, which isn't naturally available in the sequential RL data. 
This problem is circumvented by experience replay.
This method divides past experiences into single-episode updates, creating batches that are randomly sampled from a memory.
To stabilize training, two NNs are employed, a policy network, that is continuously updated and a target network that is an earlier copy of the policy network. 
The policy network estimates the current value, while the target network provides a more stable target value, represented by 
$Y$:
$$
Y_{\mathrm{DQN}}=r_{t+1}+\gamma \max_{a^{\prime}} Q_{\mathrm{target}}\left(s_{t+1}, a^{\prime}\right)
$$}

{In the Double deep Q-network (DDQN) algorithm, the action for the target value is sampled from the policy network to reduce the overestimation bias inherent in standard DQN. The corresponding target is defined as:
$$
Y_{\mathrm{DDQN}}=r_{t+1}+\gamma Q_{\mathrm{target}}\left(s_{t+1}, \underset{a^{\prime}}{\arg\max } Q_{\text {policy }}\left(s_{t+1}, a^{\prime}\right)\right) .
$$}

{This target value is approximated using a selected loss function, in this case, a smooth L$1$-norm loss.}

\section{Dependency of gates and depth on predefined threshold}\label{appndix:threshold_dependence}
{Throughout the paper, we have chosen the predefined threshold $\zeta$ constant for a fixed problem. For example, while solving two-qubit random states we choose $\zeta=10^{-5}$, which is increased to $\zeta=10^{-4}$ and later $\zeta=10^{-3}$, for the task of diagonalizing a $3$ and $4$-qubit Heisenberg model respectively. Here, we investigate the dependency of the number of gates and the depth of an RL-ansatz for a varying $\zeta$. It is straightforward to understand that the lower the $\zeta$, the more difficult it is to solve the diagonalizing problem, as a lower threshold corresponds to higher accuracy in eigenvalue estimation. Hence, we expect to observe an apparent increase in the number of gates and depth of the circuit as the threshold moves towards a lower value.}
\begin{table}[h!]
\centering
\begin{tabular}{c|cccc}
\hline
$\zeta$  & Avg. 1q gate & Avg. 2q gate & Avg. depth & Avg. num gate \\
\hline
$10^{-3}$ & 13.46        & 9.62             & 16.63      & 23.08         \\
$10^{-5}$ & 17.24        & 10.62            & 19.67      & 27.86         \\
$10^{-7}$ & 21.10        & 20.03            & 31.63      & 41.13          \\
$10^{-9}$ & 25.30        & 20.85            & 36.80       & 46.15\\ 
\hline
\end{tabular}
\caption{{\textbf{We summarize the influence of the threshold on the number of gates and depth of the RL-ansatz}. To gather data, we run 3000 episodes of RL-VQSD to solve the $3$-qubit Heisenberg model, and the results are averaged over all the successful episodes.}}
\label{tab:3q-heisen-model-threshold_influence}
\end{table}

{The results are summarized in Table~\ref{tab:3q-heisen-model-threshold_influence} where we consider the RL-VQSD to diagonalize the $3$-qubit Heisenberg model while the $\zeta$ is set from $10^{-3}$ to $10^{-9}$ in an interval of $10^{-2}$. We see that the number of gates in the circuit and the depth increase gradually as we lower the threshold.}

\section{Training time}
\label{appndix:training_time}
\begin{table}[h!]
\centering
\begin{tabular}{c|c}
\hline
CPU & \texttt{Intel(R) Core(TM) i7-10700KF CPU @ 3.80GHz}  \\
\hline
GPU & \texttt{NVIDIA GA102 [GeForce RTX 3080 Ti] 64 bits}   \\
\hline
\end{tabular}
\caption{{\textbf{The details of GPU and CPU resources utilized to record the training time}}.}
\label{tab:gpu-cpu-details}
\end{table}
{Here we discuss the time it takes to train the RL-agent in diagonalizing $2$- and $3$- qubit states. To record the time we run the RL-VQSD algorithm to diagonalize $2$-qubit arbitrary quantum state and the reduced ground state of $3$-qubit Heisenberg model for $3000$ episodes. The details of the CPU and GPU that are utilized to gather data are provided in Table~\ref{tab:gpu-cpu-details}.
To gain more insight in Table~\ref{tab:training-time-record} we compare the training time of the RL-agent with the time it takes to complete an equal amount of episodes by a random agent setting and show that in case of diagonalizing the $2$- and $3$-qubit both the methods takes the same amount of time.
\begin{table}[t!]
\centering
\begin{tabular}{c|ccc}
\hline
System & Method & Time per episode (in seconds) & Total time (in hours)  \\
\hline
\multirow{2}{3em}{2~qubit} &RL-agent & 9.21 & 11.45   \\
& Random agent & 9.29 & 11.15 \\
\hline
\multirow{2}{3em}{3~qubit} &RL-agent & 37.74 & 31.45   \\
& Random agent & 37.70 & 31.42   \\
\hline
\end{tabular}
\caption{{\textbf{The record of the training time it takes for the RL-agent and the time it takes for the random agent to complete the same number of episodes in diagonalizing a $2$ and $3$-qubit state}. As for the $2$-qubit, we choose an arbitrary full-rank state, and for $3$-qubit, we consider the reduced ground state of the Heisenberg model. The time is recorded for $3000$ episodes, which is more than sufficient to solve the diagonalization problem with the predefined threshold in the range $10^{-4}$ to $10^{-5}$.}}
\label{tab:training-time-record}
\end{table}}
\end{document}